\documentclass[prx, twocolumn, superscriptaddress, longbibliography]{revtex4-1}

\usepackage{multirow}
\usepackage[pdftex]{graphicx}
\usepackage{float}
\usepackage{color}
\usepackage{comment}

\usepackage{mathtools}
\usepackage{bm}
\usepackage{amsmath}
\usepackage{amssymb}
\usepackage{amsfonts}
\usepackage{amssymb}
\usepackage{braket}
\usepackage{amsthm}

\theoremstyle{definition}
\newtheorem{thm}{Theorem}

\newcommand{\relmiddle}[1]{\mathrel{}\middle#1\mathrel{}}
\newcommand{\integers}{\mathbb{Z}}

\newcommand{\CG}[1]{\mathop{\mathrm{CG}}(#1)}
\newcommand{\Fr}[1]{\mathop{\mathrm{Fr}}(#1)}
\newcommand{\bulk}{\mathrm{bulk}}

\newcommand{\junc}{\mathrm{junc}}

\begin{document}

\title{Exactly Solvable Spin Tri-Junctions}
\author{Masahiro Ogura}
    \email{masahiro.ogura@yukawa.kyoto-u.ac.jp}
	\affiliation{Center for Gravitational Physics and Quantum Information, Yukawa Institute for Theoretical Physics, Kyoto University, Kyoto 606-8502, Japan}
\author{Masatoshi Sato}
    \email{msato@yukawa.kyoto-u.ac.jp}
	\affiliation{Center for Gravitational Physics and Quantum Information, Yukawa Institute for Theoretical Physics, Kyoto University, Kyoto 606-8502, Japan}
\date{\today}

\begin{abstract}
We present a class of exactly solvable tri-junctions of one- and two-dimensional spin systems. Based on the geometric criterion for solvability, we clarify the sufficient condition for the junctions so that the spin Hamiltonian becomes equivalent to Majorana quadratic forms. Then we examine spin tri-junctions using the obtained solvable models. We consider the transverse magnetic field Ising spin chains and reveal how Majorana zero modes appear at the tri-junctions of the chains. Local terms of the tri-junction crucially affect the appearance of Majorna zero modes, and the tri-junction may support Majorana zero mode even if the bulk spin chains do not have Majorna end states. We also examine tri-junctions of two-dimensional SO(5)-spin lattices and discuss Majorana fermions along the junctions. 
\end{abstract}

\maketitle

\section{Introduction}
The Jordan-Wigner transformation (JWT) \cite{jordan1928pauli} maps spin operators into fermion ones. In the context of lower-dimensional exactly solvable models, the JWT has played an important role because this transformation helps solve a class of spin systems exactly \cite{onsager1944crystal,kaufman1949crystal, kaufman1949crystal2, nambu1995note, lieb1961,niemeijer1967some, katsura1962statistical, pfeuty1970one, shankar1987nearest, minami2016solvable, minami2017infinite}. For example, the JWT converts the transverse field Ising model or the one-dimensional (1D) XY model into a bi-linear of Majorana fermions, i.e., a Majorana quadratic form(MQF), enabling us to precisely calculate these models' spectra, partition functions, phase transitions, and so on. 

Whereas recent attempts have explored the generalization of fermionization in higher-dimensional spin systems \cite{kitaev2006anyons, kitaev2009topological, ryu2009three, wu2009gamma, bochniak2020bosonization, bochniak2020constraints, po2021symmetric, li2022higher, minami2019honeycomb, nussinov2009bond, cobanera2011bond, nussinov2012arbitrary, chen2019bosonization, chen2020exact, fendley2019free, chapman2020characterization, prosko2017simple, yu2008exactly, lee2007edge, shi2009topological}, the critical difference from the low-dimensional JWT is that it generally brings additional $\mathbb{Z}_2$ gauge fields. For example, a $\mathbb{Z}_2$ gauge field accompanies the MQF of the Kitaev spin model in the honeycomb lattice, which originates from redundancy in the Majorana representation of the spin variables $\mathbb{Z}_2$. Nonetheless, the model's ground state is solvable and reveals interesting quantum spin liquid phases. 

In our previous paper \cite{ogura2020geometric}, we developed a method unifying the above two techniques, which we call the single-point-connected simplicial complex (SPSC) method. The SPSC method gives a simple criterion for the solvability of lattice spin systems, based on the graph theory and the simplicial homology. Compared with the above two techniques, the SPSC method has three merits: First, it allows us to understand fermionization visually. Second, $\mathbb{Z}_2$ gauge degrees of freedom naturally emerge. Finally, this method covers many known solvable spin models and systematically gives novel ones. For example, it solves the transverse field Ising model, the XY model, and the Kitaev honeycomb spin lattice on equal footing. Moreover, we can construct new solvable spin models in higher dimensions or fractals.

In this paper, as a continuation of our previous paper, we construct exactly solvable spin junctions using the SPSC method. In particular, we examine the condition for the appearance of Majorana junction modes in exactly solvable spin tri-junctions of one- and two-dimensional spin systems.
Such junctions naturally generalize boundary problems in spin systems, and recent experiments enable their fabrication, but they have rarely been studied systematically.

The rest of this paper is organized as follows. In Sec. \ref{section_review}, we briefly review the SPSC method to construct exactly solvable models. 
In Sec. \ref{section_1djunction}, we present a class of exactly solvable tri-junctions of spin chains. We consider tri-junctions of the transverse field Ising chains and obtain the spectra. We reveal that local terms of tri-junctions significantly affect the presence of Majorana zero modes on the junctions. 
In Sec. \ref{section_2djunction}, we generalize the argument in Sec.\ref{section_1djunction} to tri-junctions of two-dimensinal $SO(5)$-spin lattices.
Finally, we give a discussion in Sec. \ref{section_discussion}.

\section{SPSC method} \label{section_review}
In this section, we summarize the main results of the SPSC method.
See Ref. \cite{ogura2020geometric} for detailed discussions and the proofs of theorems below.
For the graph theory and simplicial homology theory, see also Refs. \cite{godsil, bondy2008graph}, and \cite{kozlov2008combinatorial, nakahara}, respectively.

We start with a Hamiltonian $H$, which satisfies the following properties:

(a) $H$ has the form of $H= \sum_{j=1}^n \lambda_j h_j$ where $\lambda_j$'s are the real coefficients and $h_j$'s are operators.

(b) The operators $h_j$ satisfy
\begin{align}
    h_j^\dagger = h_j, \quad
    h_j^2=1, \quad
    h_jh_k = \eta_{j,k} h_j h_k,
\end{align}
where $\eta_{j,k}=\pm 1$.

From these operators $h_1, \ldots, h_n$, we construct the commutativity graph (CG) of the Hamiltonian $H$ as follows:

(a) Put $n$ vertices in general position and place $h_j$ on the $j$-th vertex.

(b) When $\eta_{j,k}=-1$ ($\eta_{j,k}=1$), we draw (do not draw) an edge between $j$-th and $k$-th vertices.

Below, $\CG{H}$ denotes this graph. 
Note that if $\CG{H}$ is not connected, the Hamiltonian is divided into several parts that commute with each other.
Since these parts are diagonalizable separately, 
we consider only the connected $\CG{H}$ without losing the generality.

A special class of simplicial complexes called single-point-connected simplicial complexes (SPSCs) is defined as follows.
Let $S_1, \ldots, S_m$ be simplices, and $V$ be a set consisting of all vertices in $S_1, \ldots, S_m$.
Then the set of these simplices $\{S_1, \ldots, S_m\}$ are called single-point-connected if the following conditions are satisfied:

(a) If $S_\alpha \cap S_\beta \neq \emptyset$, $S_\alpha \cap S_\beta = \{v\}$ holds for some $v \in V$.

(b) For each vertex $v \in V$, exactly two simplices in $\{S_1,\ldots,S_m\}$ include $v$.

\begin{figure}
\begin{center}
    \includegraphics[width=8cm]{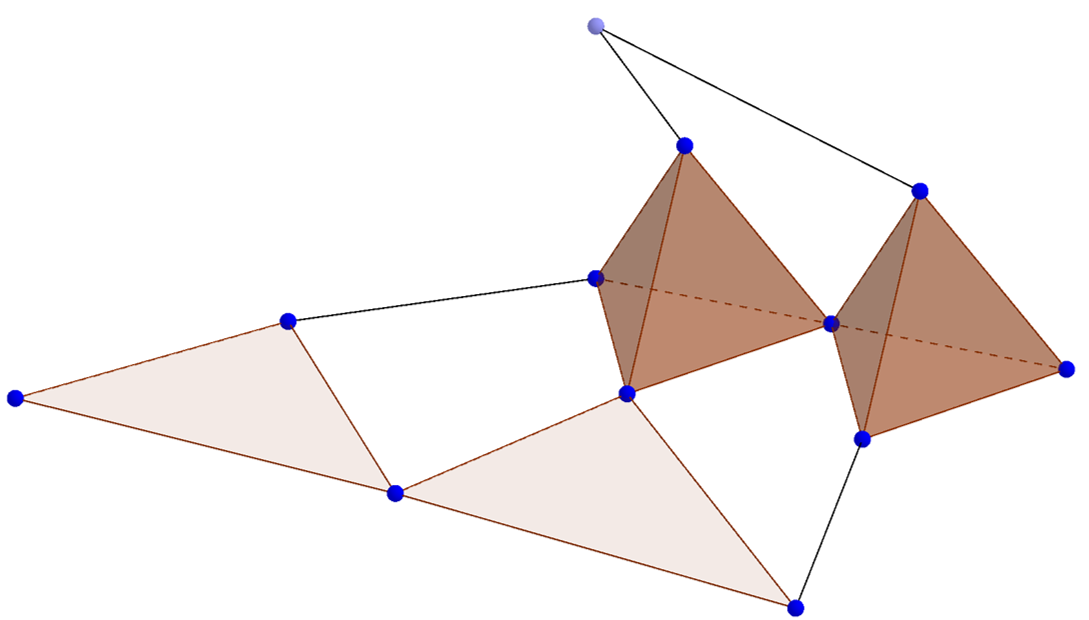}
    \caption{An example of an SPSC.
    This SPSC has $12$ vertices and $10$ generating simplices (two 3-simplices, two 2-simplices, four 1-simplices, and two 0-simplices.)
    Hence the dimension of its first homology group becomes $3$ by Theorem \ref{theorem_signs}, which corresponds to the number of holes in it.}
    \label{fig:SPSC_example}
\end{center}
\end{figure}

The following simplicial complex under this condition
\begin{align}
    K(S_1,\ldots,S_m):=\bigcup_{\alpha=1}^m \left\{ s \subset S_\alpha \relmiddle{|} \text{$s$ is a face of $S_\alpha$} \right\}
\end{align}
defines an SPSC.
An example is described in Fig. \ref{fig:SPSC_example}.
For a simplicial complex $K$, we also define the flame $\Fr{K}$ of $K$ as a graph whose vertices and edges are $0$-faces and $1$-faces in $K$, respectively.

Under the above preparation, we introduce an SPSC Hamiltonian.
We call the Hamiltonian $H$ an SPSC Hamiltonian for $K$, if there exists an SPSC $K=K(S_1, \ldots, S_m)$ obeying $\Fr{K}=\CG{H}$.
Now we explain the main theorem:

\begin{thm} \label{theorem_transformation}
Let $H= \sum_{j=1}^n \lambda_j h_j$ be an SPSC Hamiltonian for a connected SPSC $K=K(S_1,\ldots,S_m)$.
Then $H$ is mapped to a quadratic form of $m$ Majorana operators $\varphi_1, \ldots, \varphi_m$ that are put on the simplices $S_1,\ldots, S_m$, respectively.
More precisely, if the $j$-th vertex is shared by $S_\alpha$ and $S_\beta$, $h_j$ is described as
\begin{align} \label{into_MQF}
    h_j= -i \epsilon_j \varphi_\alpha \varphi_\beta,
\end{align}
where $\epsilon_j = \pm 1$.
\end{thm}

The sign factors $\epsilon_j$ above can be regarded as a $\mathbb{Z}_2$ gauge field and are determined as follows.
Using the $\mathbb{Z}_2$ gauge transformation $\varphi_\alpha\to - \varphi_\alpha$, we can change the sign factor $\epsilon_j$ in Eq. (\ref{into_MQF}) without changing (anti-)commutation relations of $h_j$s.
While this procedure trivializes $m-1$ relative signs,  $n-m+1$ sign factors remain unfixed.
For the unfixed sign factors, we have the next theorem. 
\footnote{We also present another theorem for the conserved quantities. See Ref.\cite{ogura2020geometric}. }.
\begin{thm} \label{theorem_signs}
Under the same assumption of Theorem \ref{theorem_transformation},
\begin{align}
    \dim H_1(K; \mathbb{Z}_2) = n-m+1
\end{align}
holds, i.e., there exist $n-m+1$ holes in $K$.
Correspondingly, we obtain $n-m+1$ conserved quantities from $h_j$s on these holes, determining the unfixed sign factors. 
\end{thm}

From these two theorems, we can solve various spin systems exactly in terms of free fermion systems with $\mathbb{Z}_2$ gauge fields $\epsilon_j$ \cite{ogura2020geometric}.

\section{Tri-Junctions of 1D spin chains} \label{section_1djunction}
In this section, we construct exactly solvable tri-junctions of one-dimensional (1D) spin chains and examine their properties.

\subsection{Transverse Field Ising Model}
\label{sec:tfim}
\begin{figure}[tbp]
    \centering
    \includegraphics[width=8cm]{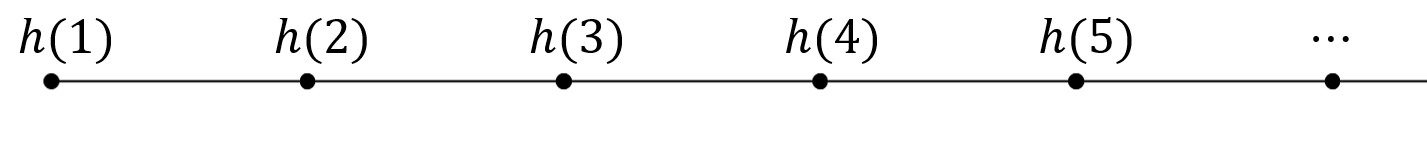}
    \caption{The CG of the transverse field Ising chain.}
    \label{fig_TFIM_CG}
\end{figure}

Before discussing junction models, we solve the transverse field Ising chain by the SPSC method.
The Hamiltonian of the transverse field Ising chain reads
\begin{align} \label{TFIM_hamiltonian}
    H=
    -J\sum_{j=1}^\infty \sigma^z(j) \sigma^z(j+1)
    -h \sum_{j=2} ^\infty \sigma^x(j)
    -t \sigma^x(1),
\end{align}
where $\sigma^z(j)$ and $\sigma^x(j)$ are the $z$- and $x$-components of the Pauli matrices at site $j$, $J$ is the exchange constant, $h$ is the transverse magnetic field, and $t$ is a magnetic field at the endpoint of the chain.
Assigning $h(j)$ as $h(1)=\sigma^z(1)$ and
\begin{align}
    h(2j)= \sigma^z(j) \sigma^x(j+1),\quad
    h(2j+1)=\sigma^x(j+1),
    \label{eq:htfi}
\end{align}
for $j=1,2,\dots$, we have the CG in Fig. \ref{fig_TFIM_CG}.
Regarding each edge and vertex in the CG as a 1-simplex and a 0-simplex, we can identify the CG as an SPSC.
Since there is no hole in the CG,
the corresponding Hamiltonian with the Majorana operator $\varphi(j)$ reads
\begin{align} \label{TFIM_transformed}
    H=
    &it\varphi (0) \varphi (1)
    +iJ\sum_{j=1}^\infty \varphi(2j-1) \varphi (2j) \nonumber \\
    &+ih\sum_{j=1}^\infty \varphi (2j) \varphi (2j+1),
\end{align}
where we fix the $\integers_2$ gauge field as $+1$ by 
the gauge transformation.
We represent this Hamiltonian in the matrix form,
\begin{align}
    H=\frac{i}{2} \Vec{\varphi}^\dagger \mathcal{A} \Vec{\varphi}, 
\end{align}
where
\begin{align} \label{TFIM_matrix}
    \Vec{\varphi}&=
    \begin{pmatrix}
        \varphi(0) & \varphi(1) & \varphi(2) & \varphi(3) & \cdots
    \end{pmatrix}^T, \nonumber \\
    \mathcal{A}&=
    \begin{pmatrix}
        0  & t  &    &   &   &\\
        -t & 0  & J  &   &   &\\
           & -J & 0  & h &   &\\
           &    & -h & 0 & \ddots &\\
           &    &    & \ddots & \ddots
    \end{pmatrix}. 
\end{align}
For an eigenvalue $\lambda$ of $\mathcal{A}$ with an eigenvector
\begin{align}
    \Vec{v}=
    \begin{pmatrix}
        v(0) & w(0) & v(1) & w(1) & \cdots
    \end{pmatrix}^T,
\end{align}
the eigenequation is given by
\begin{align}
    tw(0) &= \lambda v(0), \label{TFIM_eigen_1}\\
    -tv(0) + Jv(1) &= \lambda w(0), \label{TFIM_eigen_2} \\
    -Jw(n-1) +hw(n) &= \lambda v(n), \label{TFIM_eigen_3} \\
    -hv(n) + Jv(n+1) &=\lambda w(n), \label{TFIM_eigen_4}
\end{align}
with $n=1,2,\ldots$.
Note that the eigenvalue $\lambda$ gives an energy $E$ of the system by $E=i\lambda$.

To solve the eigenequation, we define the generating functions:
\begin{align}
    V(z)=\sum_{n=1}^\infty v(n)z^n, \quad
    W(z)=\sum_{n=1}^\infty w(n)z^n.
\end{align}
Then from Eqs. (\ref{TFIM_eigen_2}), (\ref{TFIM_eigen_3}) and (\ref{TFIM_eigen_4}), we get
\begin{align}
    \begin{pmatrix}
        \lambda & Jz-h \\
        h-Jz^{-1} & \lambda 
    \end{pmatrix}
    &\begin{pmatrix}
        V(z) \\
        W(z)
    \end{pmatrix} \nonumber \\
    =v(0)
    &\begin{pmatrix}
        0 \\
        -t
    \end{pmatrix}
    +w(0)
    \begin{pmatrix}
        -Jz \\
        -\lambda
    \end{pmatrix}.
\end{align}
Thus, $V(z)$ and $W(z)$ are given by
\begin{align} \label{TFIM_mother_calculated}
    V(z)
    =&\frac{1}{hJ}\frac{z}{z^2-2 \kappa z+1} \nonumber \\
    &\times \left[t \left(h-Jz \right)v(0)+ h\lambda w(0) \right], \nonumber \\
    W(z)
    =&\frac{1}{hJ}\frac{z}{z^2-2 \kappa z+1} \nonumber \\
    &\times \left[t \lambda v(0)-(J^2+\lambda^2-hJz) w(0) \right],
\end{align}
where
\begin{align}
    \kappa:=\frac{J^2+h^2+ \lambda^2}{2Jh}.
\end{align}

Now we examine the spectrum of this Hamiltonian.
Let $z_\pm$ be the roots of $z^2-2\kappa z +1$. Since they satisfy
\begin{align} \label{TFIM_rootandcoefficients}
    z_++z_-=2\kappa, \quad
    z_+ z_- =1,
\end{align}
we have
\begin{align}
    \frac{z}{z^2-2\kappa z +1}
    &=\frac{1}{z_+-z_-}\left(\frac{z_+z}{1-z_+z}-\frac{z_-z}{1-z_-z}\right) \nonumber \\
    &=\sum_{n=1}^\infty \left( \frac{z_+^n-z_-^n}{z_+-z_-} \right) z^n.
\end{align}
We require $|z_\pm|=1$ in order to obtain the bulk spectrum.
Setting $z_\pm = e^{ik}$ with $k \in (-\pi,\pi]$, then from Eq. (\ref{TFIM_rootandcoefficients}), we obtain the bulk spectrum as
\begin{align}
    E = \pm \sqrt{J^2+h^2-2Jh \cos k} \quad (k \in (-\pi,\pi]),
\end{align}
which reproduces the result in Ref. \cite{minami2016solvable}.
Remark that the bulk spectrum becomes gapless if $J=h$ holds.

In contrast, for $|z_\pm| \neq 1$, we have boundary modes.
We focus on zero energy boundary modes with $\lambda=0$,
then from Eq. (\ref{TFIM_mother_calculated}), we obtain
\begin{align} \label{TFIM_edge_mother}
    V(z)
    =\frac{tz}{J-hz}v(0)
    =\frac{t}{h}v(0) \sum_{n=1}^\infty \left( \frac{hz}{J} \right)^n, \nonumber \\
    W(z)
    =\frac{Jz}{h-Jz}w(0)
    =w(0) \sum_{n=1}^\infty \left( \frac{Jz}{h} \right)^n.
\end{align}
Moreover, we also get $ w(0)=0$ from Eq. (\ref{TFIM_eigen_1}), which implies $W(z)=0$.
So the boundary state becomes
\begin{align}
    v(0) \cdot
    \begin{pmatrix}
        1 & 0 & \frac{t}{h} \cdot \frac{h}{J} & 0 & \frac{t}{h} \cdot \left( \frac{h}{J} \right)^2 & 0 & \cdots
    \end{pmatrix}^T,
\end{align}
which satisfies the $L^2$ condition for $h<J$.
Therefore, under this condition, a zero energy boundary state appears.

In summary, we find a phase transition at $J=h$, and the transverse field Ising chain supports a zero energy boundary state for $h<J$.

\subsection{Exactly Solvable Spin Tri-junctions}
\label{sec:stj1d}
\begin{figure}
    \centering
    \includegraphics[width=8cm]{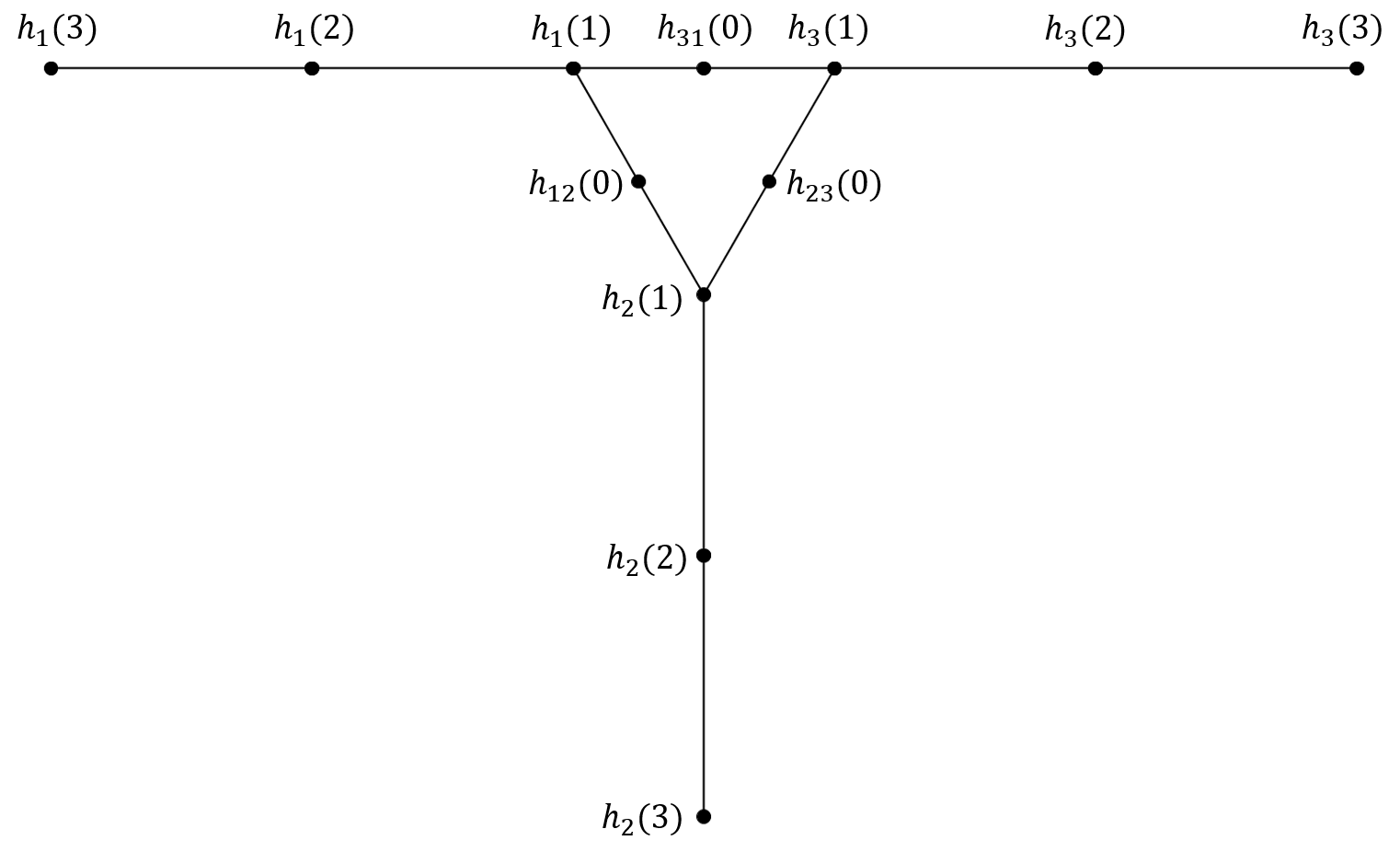}
    \caption{The CG of the tri-junction in Eq.(\ref{eq:tj}).
    This CG is not an SPSC.}
    \label{fig:1Djunction_typical}
\end{figure}

In this subsection, we discuss patterns of spin tri-junctions considered in this paper.
We consider tri-junctions of the transverse field Ising chains, of which Hamiltonian is given by
\begin{align} \label{1djunction_hamiltonian}
     H&=\sum_{\mu=1}^3 H_\mu^\bulk + H^\junc .
\end{align}
Here $H_{\mu}^\bulk$ is the semi-infinite bulk spin chain Hamiltonian
\begin{align}
    &H_\mu^\bulk \nonumber \\
    &=- \sum_{\mu=1}^3 \left[ J_\mu \sum_{j=1}^\infty \sigma_\mu^z(j) \sigma_\mu^z(j+1) + h_\mu \sum_{j=2}^\infty \sigma_\mu^x(j) \right],
\end{align}
with the index $\mu=1,2,3$ specifying the chains.
Like Eq. (\ref{eq:htfi}), we assign $h_\mu(j)$ as
\begin{align}
&h_\mu(2j)=\sigma_\mu^z(j) \sigma_\mu^z(j+1),\quad 
h_\mu(2j+1)=\sigma_\mu^x(j+1), 
\nonumber\\
&(j=1,2,\dots).
\end{align}
Below, we determine the junction part $H^\junc$ so that the CG of $H$ becomes an SPSC.
We do not treat a junction
\begin{align}
    H^\junc=
    &-\sum_{\mu=1}^3 h_\mu \sigma_\mu^x(1) -g_{12} \sigma_1^z(1) \sigma_2^z(1) \nonumber \\
    &-g_{23} \sigma_2^z(1) \sigma_3^z(1)
    -g_{31} \sigma_3^z(1) \sigma_1^z(1),
\label{eq:tj}
\end{align}
with the interchain couplings $g_{\mu\nu}$ \cite{tsvelik2013majorana,giuliano2016junction,giuliano2020emerging}.
From the assignment,
$
h_{\mu,\nu}(0)=\sigma_\mu^z(1) \sigma_\nu^z(1),
$
we obtain the CG in Fig. \ref{fig:1Djunction_typical}, but this is not an SPSC. 

\begin{figure*}[tbp]
    \begin{minipage}{0.32\linewidth}
        \centering
        \includegraphics[width=55mm]{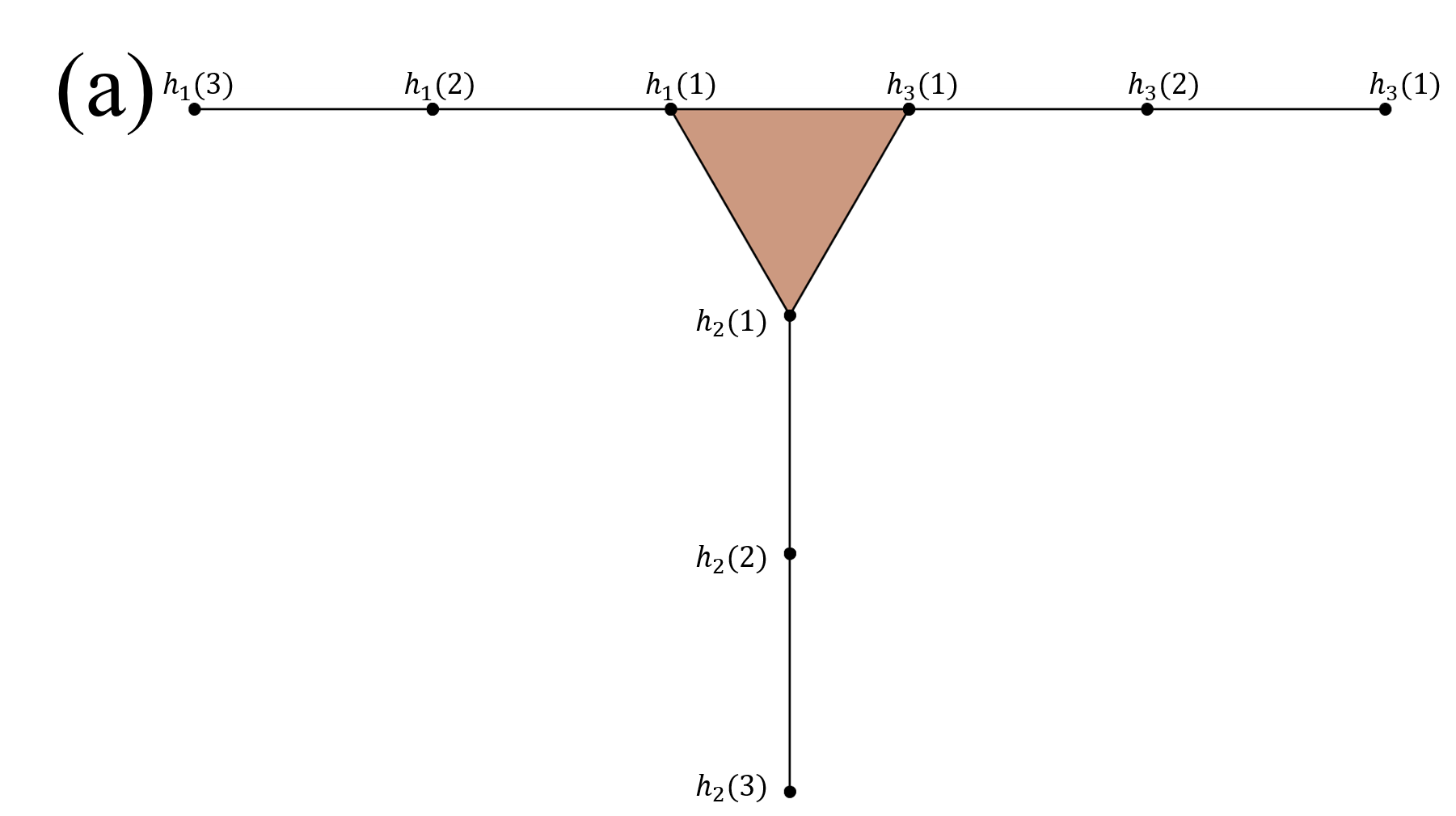}
    \end{minipage} 
    \begin{minipage}{0.32\linewidth}
        \centering
        \includegraphics[width=55mm]{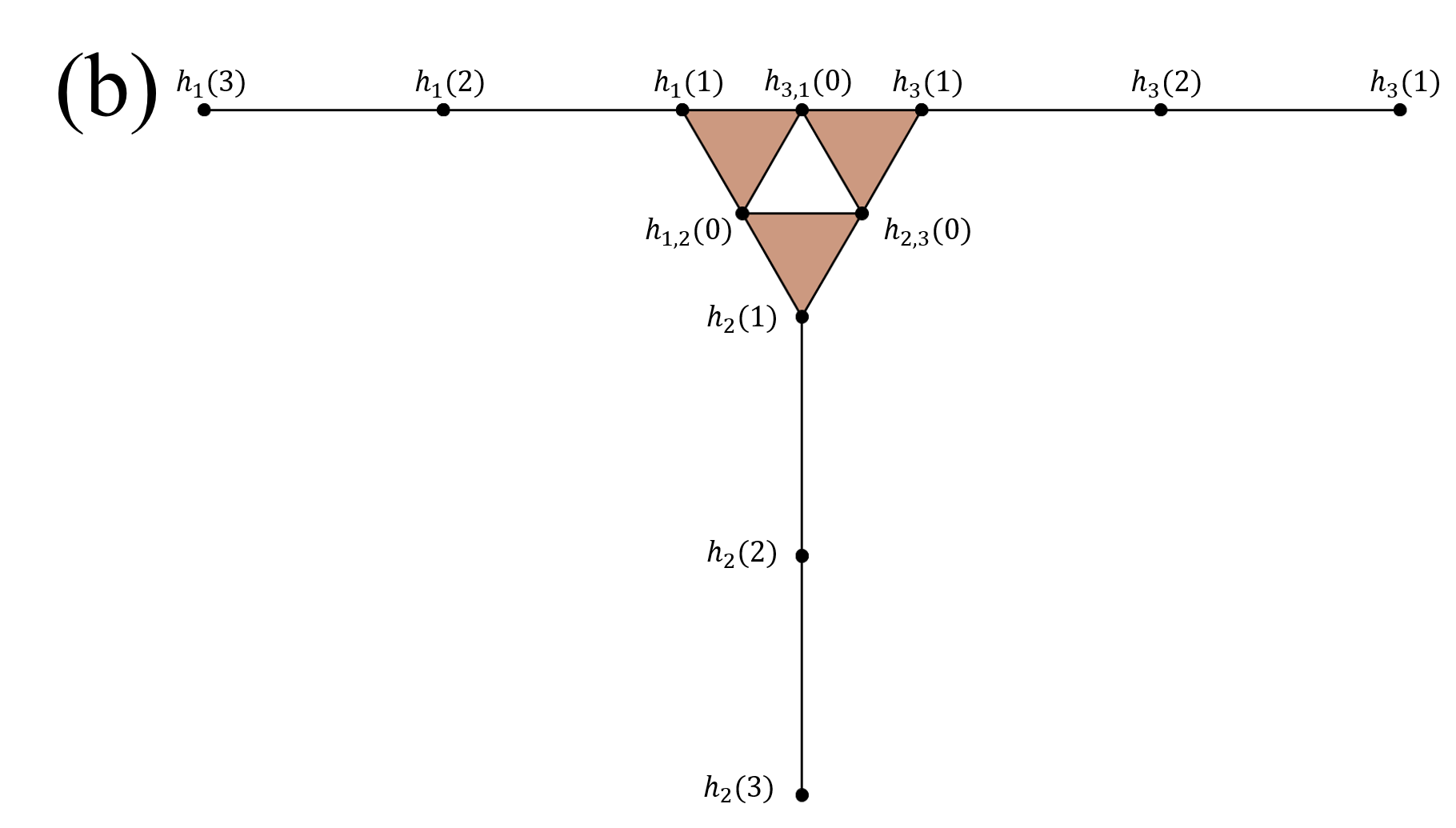}
    \end{minipage} 
    \begin{minipage}{0.32\linewidth}
        \centering
        \includegraphics[width=55mm]{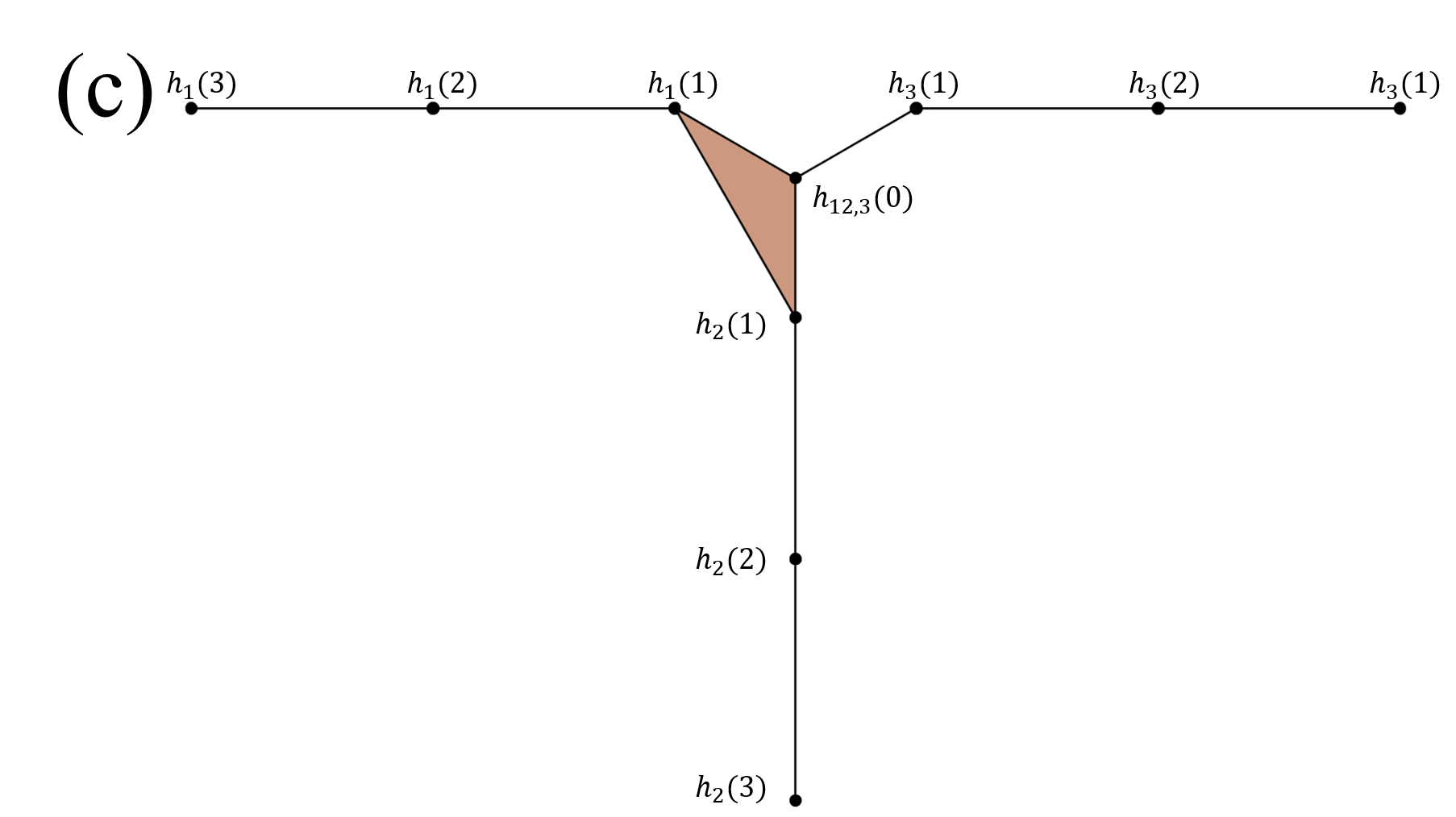}
    \end{minipage}
    \caption{Three patterns of exactly solvable tri-junctions.(a) $(3)$-tri-junction. (b) $(1+1+1)$-tri-junction. (c) $(2+1)$-tri-junction.}
    \label{figs:1Djunctions}
\end{figure*}

We find that the above requirement for $H^\junc$ is too general, allowing an infinite number of patterns of tri-junctions.
To avoid the artificial complexity, we restrict the form of the junction as follows:
\begin{align}
    H^\junc=
    -\sum_{\mu=1}^3 t_\mu h_\mu(1)
    - \sum_{\omega, \omega' \in \Omega, \omega \neq \omega'} g_{\omega,\omega'} h_{\omega,\omega'}(0),
\label{eq:junc}
\end{align}
where $\Omega$ is a set whose elements $\omega$ are subsets of $\{1,2,3\}$ and satisfy $\bigsqcup_{\omega \in \Omega} \omega = \{1,2,3\}$.
More explicitly, $\Omega$ is one of the following sets
\begin{align} \label{omega_patterns}
    \{\{123\}\}, \quad
    \{\{12\},\{3\}\}, \quad
    \{\{1\},\{2\},\{3\}\},
\end{align}
and their permutations of the numbers.
The first term in $H^\junc$ is the coupling to the spin chains, and
the second term is the intra-junction coupling commuting with $H^\bulk$.
We assume the simple coupling to the spin chains, $\{h_\mu(1), h_\mu(2)\}=0$, which satisfies the locality of the interaction and the connectivity in the resultant CG. 
This assumption also allows the mapping of the bulk spin chains to MQFs similar to Eq.  (\ref{TFIM_transformed}):
\begin{align}
    H_\mu^\bulk=i\sum_{j=1}^\infty &\left[ J_\mu \varphi_\mu (2j-1) \varphi_\mu (2j) \right. \nonumber \\
    &\left. +h_\mu \varphi_\mu (2j) \varphi_\mu (2j+1)
    \right].
\end{align}
For each $\Omega$, 
we specify the intra-junction coupling as follows:
(i) $h_\mu(1)$ and $h_\nu(1)$ anti-commute with each other if there exists $\omega \in \Omega$ such that $\mu,\nu \in \omega$.
(ii) $h_\mu(1)$ anti-commutes with $h_{\omega,\omega'}(0)$ when $\mu \in \omega \cup \omega'$.
(iii) $h_{\omega, \omega'}(0)$ and $h_{\omega'',\omega'''}(0)$ anti-commutes with each other if they share a single subindex $\omega$.
In other cases, we assume the commutation relations.
The corresponding CGs show three patterns of SPSCs, as illustrated in Fig. \ref{figs:1Djunctions}.

These patterns can be understood as follows.
Let $p(n)$ be the partition of a positive integer $n$, which is the number of possible divisions of $n$ into positive integers.
Then, we have $p(3)=3$ as $3$ can be decomposed into positive integers in three different ways,
\begin{align}
    3
    &=3 \nonumber \\
    &=2+1 \nonumber \\
    &=1+1+1.
\end{align}
Because these divisions correspond to three patterns of tri-junctions,
we call the tri-junctions in Fig \ref{figs:1Djunctions} as
$(3)$-, $(2+1)$-, and $(1+1+1)$-tri-junctions, respectively.

Similarly, we can easily obtain exactly solvable models for higher multi-junctions by using $H^\junc$ similar to Eq. (\ref{eq:junc}).
For example,  because it holds $p(4)=5$,
\begin{align}
    4
    &=4 \nonumber \\
    &=3+1 \nonumber \\
    &=2+2 \nonumber \\
    &=2+1+1 \nonumber \\
    &=1+1+1+1,
\end{align}
we have five different patterns of exactly solvable tetra-junctions.

\subsection{Majorana zero modes in tri-junctions}
Here we examine various exactly solvable tri-junctions in Eq. (\ref{1djunction_hamiltonian}).
Since each bulk Hamiltonian $H_\mu^\bulk$ is essentially identical to Eq. (\ref{TFIM_hamiltonian}),
the bulk eigenequation for the spin chains is essentially the same as Eqs. (\ref{TFIM_eigen_3}) and (\ref{TFIM_eigen_4}),
\begin{align}
    -J_\mu w_\mu(n-1)+hw_\mu (n) &=\lambda v_\mu(n), \label{1djunction_eigen_bulk_1} \\
    -h_\mu v_\mu(n) + J_\mu v_\mu(n+1) &=\lambda w_\mu(n), \label{1djunction_eigen_bulk_2}
\end{align}
with $n=1,2,\ldots$.
Thus, in a manner similar to Sec.\ref{sec:tfim}, we can analyze the junctions by introducing
the generating functions
\begin{align} \label{1djunction_mother_def}
    V_\mu(z):=\sum_{n=1}^\infty v_\mu(n)z^n, \quad
    W_\mu(z):=\sum_{n=1}^\infty w_\mu(n)z^n,
\end{align}
The eigenequation for the junction part provides a boundary condition for the generating functions.

\begin{table*}[htbp]
    \centering
    \caption{
    The number of Majorana zero modes localized at the 1D $(3)$-, $(1+1+1)$-, and $(2+1)$-tri-junctions.
    }
    \begin{tabular}{c|cccc} \hline \hline
        Conditions & $(3)$-tri-junction (I) & $(3)$-tri-junction (II) & $(1+1+1)$-tri-junction & $(2+1)$-tri-junction  \\ \hline
        $h_1<J_1, h_2<J_2, h_3<J_3$ & 1 & 2 & 1 & 0 \\
        $h_1>J_1, h_2<J_2, h_3<J_3$ & 0 & 1 & 0 & 1 \\
        $h_1<J_1, h_2<J_2, h_3>J_3$ & 0 & 1 & 0 & 1 \\
        $h_1>J_1, h_2>J_2, h_3<J_3$ & 1 & 0 & 1 & 2 \\
        $h_1>J_1, h_2<J_2, h_3>J_3$ & 1 & 0 & 1 & 0 \\
        $h_1>J_1, h_2>J_2, h_3>J_3$ & 2 & 1 & 0 & 1 \\ \hline \hline
    \end{tabular}
    \label{table_edgemode}
\end{table*}

\subsubsection{$(3)$-Tri-Junction (I)} \label{section_1djunction_(3)}
In the next two subsections,  we analyse two different models for the $(3)$-tri-junction.
First, in this subsection, we consider the following junction:
\begin{align} 
    H^\junc=
    &-t_1 \sigma_1^x(1) \sigma_2^z(1) 
    -t_2 \sigma_2^x(1) \sigma_3^z(1) \nonumber \\
    &-t_3 \sigma_3^x(1) \sigma_1^z(3).
\label{eq:(3)I}
\end{align}
Defining $h_\mu(1)$ as
\begin{align}
&h_1(1)=\sigma_1^x(1)\sigma_2^z(1), \quad    h_2(1)=\sigma_2^x(1)\sigma_3^z(1), 
\nonumber\\
&h_3(1)=\sigma_3^x(1)\sigma_1^z(1),    
\end{align}
we have the CG in Fig. \ref{figs:1Djunctions} (a).
This CG is an SPSC by regarding each edge as a 1-simplex and the central triangle as a 2-simplex. 
As a result, we can map the Hamiltonian into a MQF:
\begin{align} \label{(3)junction_MQF}
    H^\junc=
    i \sum_{\mu=1}^3 t_\mu \varphi_{123}(0) \varphi_\mu(1),
\end{align}
where $\varphi_{123}(0)$ is a Majorana operator on the triangle.
The corresponding eigenequation for the junction part is given by
\begin{align}
    \sum_{\mu=1}^3 t_\mu w_\mu(0) &= \lambda v_{123}(0), \label{1djunction_eigen_(3)edge_1} \\
    -t_\mu v_{123}(0) + J_\mu v_\mu (1) &= \lambda w_\mu(0) \quad (\mu=1,2,3). \label{1djunction_eigen_(3)edge_2}
\end{align}
From Eqs. (\ref{1djunction_eigen_bulk_1}), (\ref{1djunction_eigen_bulk_2}), and (\ref{1djunction_eigen_(3)edge_2}), the generating functions in Eq. (\ref{1djunction_mother_def})
are determined as
\begin{align} \label{1djunction_bulk_mother_nonzero}
    V_\mu(z)
    =&\frac{1}{h_\mu J_\mu}\frac{z}{z^2-2 \kappa_\mu z+1} \nonumber \\
    &\times \left[t_\mu \left(h_\mu - J_\mu z \right)v_{123}(0)+h_\mu \lambda w_\mu(0) \right], \nonumber \\
    W_\mu(z)
    =&\frac{1}{h_\mu J_\mu}\frac{z}{z^2-2 \kappa_\mu z+1} \nonumber \\
    &\times \left[t_\mu \lambda v_{123}(0)-(J_\mu^2+\lambda^2-h_\mu J_\mu z) w_\mu(0) \right],
\end{align}
with
\begin{align}
    \kappa_\mu:=
    \frac{J_\mu^2+ h_\mu ^2 + \lambda^2}{2J_\mu h_\mu}.
\end{align}

Now we focus on zero modes with $\lambda=0$.
For zero modes, the generating functions read
\begin{align} \label{1djunction_edge_(3)_mother}
    V_\mu(z)
    &=\frac{t_\mu}{h_\mu}v_{123}(0) \sum_{n=1}^\infty \left( \frac{h_\mu z}{J_\mu} \right)^n, \nonumber \\
    W_\mu(z)
    &=w_\mu(0) \sum_{n=1}^\infty \left( \frac{J_\mu z}{h_\mu} \right)^n,
\end{align}
where $w_\mu(0)$ obeys Eq.(\ref{1djunction_eigen_(3)edge_1}) with $\lambda=0$, 
\begin{align} \label{1djunction_edge_(3)_relation}
    \sum_{\mu=1}^3 t_\mu w_\mu(0)=0.
\end{align}
To keep the $L^2$ condition of the corresponding eigenvectors, $v_\mu(n)$ and $w_\mu(n)$ must go to zero when $n\to\infty$.  
Thus, from Eq. (\ref{1djunction_edge_(3)_mother}), we have $w_\mu(0)=0$ ($v_{123}(0)=0$) if $h_\mu<J_\mu$ ($h_\mu>J_\mu$).
Then, we have four different situations: (i) All the spin chains satisfy $h_\mu<J_\mu$. (ii) One of the spin chains, say the first one satisfies $h_1>J_1$, and the other chains satisfy $h_2<J_2$ and $h_3<J_3$. (iii) Two of the spin chains, say the first and the second ones satisfy $h_1>J_1$, and $h_2>J_2$, and the other satisfies $h_3<J_3$. (iv) All the spin chains satisfy $h_\mu>J_\mu$.
For each of these cases, we can count the number of zero modes as follows.

For case (i), we have $w_1(0)=w_2(0)=w_3(0)$, which obeys Eq. (\ref{1djunction_edge_(3)_relation}). Thus, we have a single zero mode generated by $V_\mu(z)$ in Eq. (\ref{1djunction_edge_(3)_mother}) with $v_{123}(0)\neq 0$. For case (ii), we have $w_2(0)=w_3(0)$ and $v_{123}(0)=0$ from the $L^2$ condition. From Eq. (\ref{1djunction_edge_(3)_relation}), we also have $w_3(0)=0$. Hence $V_\mu(z)$ and $W_\mu(z)$
are identically zero, and thus no zero mode exists.  
For case (iii), the $L^2$ condition and Eq. (\ref{1djunction_edge_(3)_relation}) lead to $w_3(0)=v_{123}(0)=0$ and $t_1 w_1(0) + t_2 w_2(0)=0$. Thus, we have a single independent zero mode generated by $W_1(z)$ and $W_2(z)$ with $t_1 w_1(0) + t_2 w_2(0)=0$.
Finally, for case (iv), a similar argument gives $v_{123}(0)=0$ and $t_1 w_1(0) + t_2 w_2(0) + t_3 w_3(0)=0$, which lead to two independent zero modes generated by $W_{\mu=1,2,3}(z)$.
All the obtained zero modes above are Majorana fermions localized at the junction. 
We summarize the number of Majorana junction modes in Table \ref{table_edgemode}.

\subsubsection{$(3)$-Tri-Junction (II)}
Here we examine a different model for the $(3)$-tri-junction.
The junction Hamiltonian is given by
\begin{align}
    H^\junc=
    &-t_1 \sigma_1^x(1) -t_2 \sigma_2^x(1) -t_3 \sigma_3^x(1) \nonumber \\
    &-g_1 \sigma_1^z(1) S^x -g_2 \sigma_2^z(1) S^y -g_3 \sigma_3^z(1) S^z,
\end{align}
where $S^{a}$ $(a=x,y,z)$ is the Pauli matrix for an additional spin on the junction.
Whereas the CG of this junction has the same form as Fig. \ref{figs:1Djunctions}(a),
the properties of the junction are very different from those in Eq. (\ref{eq:(3)I}) 
as we will show below.

Since this CG is an SPSC,
the junction Hamiltonian can be rewritten in the form of MQFs by using the SPSC method,
\begin{align} \label{(3)junctionII-hamiltonian}
    H^\junc=
    i \sum_{\mu=1}^3 \left[ t_\mu \varphi_\mu(0) \varphi_\mu(1)
    + g_\mu \varphi_{123}(0) \varphi_\mu(0) \right],
\end{align}
where $\varphi_\mu(0)$ and $\varphi_{123}(0)$ are Majorana operators.
The corresponding junction part of the eigenequation is
\begin{align}
    \sum_{\mu=1}^3 g_\mu v_\mu (0) &= \lambda v_{123}(0), \nonumber \\
    -g_\mu v_{123}(0) +t_\mu w_\mu(0) &= \lambda v_\mu (0), \nonumber \\
    -t_\mu v_\mu(0) + J_\mu v_\mu(1) &= \lambda w_\mu(0),
\end{align}
with $\mu=1,2,3$.
Then, we find that the generating functions in Eq. (\ref{1djunction_mother_def}) for zero modes with $\lambda=0$ become
\begin{align} \label{1djunction_edge_(3)II_mother}
    V_\mu(z)
    &=\frac{t_\mu}{h_\mu}v_{\mu}(0) \sum_{n=1}^\infty \left( \frac{h_\mu z}{J_\mu} \right)^n, \nonumber \\
    W_\mu(z)
    &=\frac{g_\mu}{t_\mu} v_{123}(0) \sum_{n=1}^\infty \left( \frac{J_\mu z}{h_\mu} \right)^n,
\end{align}
with
\begin{align}
    \sum_{\mu=1}^3 g_\mu v_\mu (0) = 0.
\end{align}

In a manner similar to Sec. \ref{section_1djunction_(3)}, we can count the number of Majorana junction modes by demanding the $L^2$ condition. 
Note that the role of $V_\mu(z)$ and $W_\mu(z)$ is interchanged between Eqs.(\ref{1djunction_edge_(3)_mother}) and (\ref{1djunction_edge_(3)II_mother}).
Thus, the number of Majorana zero modes in the present junction does not coincide with that in the previous subsection.
In particular, the parity of the Majorana mode number is opposite between these junctions.
We summarize the obtained result in Table \ref{table_edgemode}.

\subsubsection{$(1+1+1)$-Tri-Junction}
\label{sec:(1+1+1)1D}
For a model of the $(1+1+1)$-tri-junction, we consider
\begin{align}
    H^\junc=
    &-t_1 \sigma_1^x(1) -t_2 \sigma_2^x(1) -t_3 \sigma_3^x(1) \nonumber \\
    &-g_{12} \sigma_1^z(1)\sigma_2^z(1)S^x
    -g_{23} \sigma_2^z(1)\sigma_3^z(1)S^y \nonumber \\
    &-g_{31} \sigma_3^z(1)\sigma_1^z(1)S^z,
\end{align}
where $S^a$ is the Pauli matrix for an additional spin.
The CG of this Hamiltonian is an SPSC in Fig. \ref{figs:1Djunctions} (b). The Hamiltonian in an MQF is
\begin{align}
    H^\junc=&i \sum_{\mu=1}^3 t_\mu \varphi_\mu(0) \varphi_\mu(1) +ig_{12} \epsilon_{12} \varphi_{1}(0) \varphi_2(0) \nonumber \\
    &+ig_{23} \epsilon_{23} \varphi_{2}(0) \varphi_3(0)
    +ig_{31} \epsilon_{31} \varphi_{3}(0) \varphi_1(0),
\end{align}
with $\epsilon_{\mu\nu} = \pm 1$.
Here using the $\mathbb{Z}_2$ gauge transformation, we have set $\epsilon_{12}=\epsilon_{23}=1$.
The remaining sign $\epsilon_{13}$ is determined by the conserved quantity of the hole in Fig. \ref{figs:1Djunctions} (b).
In the present case, the conserved quantity is $i$, which implies $\epsilon_{31}=1$.

The eigenequation for the junction is
\begin{align}
    t_1 w_1(0) + g_{12}v_2(0) -g_{31} v_3(0) &= \lambda v_1(0), \nonumber \\
    -g_{12}v_1(0) + t_2w_2(0) + g_{23} v_3(0) &= \lambda v_2(0), \nonumber \\
    g_{31} v_1(0) - g_{23}v_2(0) + t_3 w_3(0) &= \lambda v_3(0), \nonumber \\
    -t_\mu v_\mu(0) +J_\mu v_\mu(1) &= \lambda w_\mu(0),
\label{eq:(1+1+1)-1Djunc}
\end{align}
with $\mu=1,2,3$.
Equations. (\ref{1djunction_eigen_bulk_1}) and (\ref{1djunction_eigen_bulk_2}) still work as the bulk eigenequation for the spin chains, and hence we have the bulk generating functions in the form of Eq. (\ref{1djunction_bulk_mother_nonzero}) with replacing $v_{123}(0)$ by $v_\mu(0)$.

For zero modes with $\lambda=0$, we have Eq. (\ref{1djunction_edge_(3)_mother}) with replacing $v_{123}(0)$ by $v_\mu(0)$. We also have the constraint at the junction
\begin{align}
    t_1 w_1(0) + g_{12}v_2(0) - g_{31} v_3(0) &=0, \nonumber \\
    -g_{12}v_1(0) + t_2w_2(0) + g_{23} v_3(0) &= 0, \nonumber \\
    g_{31} v_1(0) - g_{23}v_2(0) + t_3 w_3(0) &=0.
\end{align}
Considering the $L^2$ condition, we can count  Majorana junction modes. 
The result is summarized in Table \ref{table_edgemode}.

\subsubsection{$(2+1)$-Tri-Junction}
An example of the $(2+1)$-tri-junction is
\begin{align}
    H^\junc=&-t_1 \sigma_1^x(1)
    -t_2 \sigma_2^x(1) \sigma_1^z(1)
    -t_3 \sigma_3^x(1) \nonumber \\
    &-g_{12,3} \sigma_1^z(1)\sigma_2^z(1)\sigma_3^z(1).
\end{align}
The CG of this model is  Fig. \ref{figs:1Djunctions}(c), which is an SPSC.
The junction Hamiltonian in the form of an MQF is,
\begin{align}
    H^\junc=
    &i t_1 \varphi_{12}(0) \varphi_1(1)
    +i t_2 \varphi_{12}(0) \varphi_2(1) \nonumber \\
    &+i t_3 \varphi_{3}(0) \varphi_3(1)
    +i g_{12,3} \varphi_{12}(0) \varphi_3(0).
\end{align}
Hence the eigenequation for the junction becomes
\begin{align}
    t_1 w_1(0) + t_2 w_2(0) + g_{12,3} v_3(0) &= \lambda v_{12}(0), \nonumber \\
    -g_{12,3} v_{12}(0) + t_3 w_3(0) &= \lambda v_3(0), \nonumber \\
    -t_\mu v_\mu(0) + J_\mu v_\mu (1) &= \lambda w_\mu(0),
\end{align}
with $\mu=1,2,3$.
For zero modes with $\lambda = 0$, we have again
Eq. (\ref{1djunction_edge_(3)_mother}) with replacing $v_{123}(0)$ by $v_{12}(0)$ ($v_3(0)$) in $V_{\mu=1,2}(z)$ ($V_3(z)$).
We also have the boundary condition at the junction,
\begin{align}
    t_1 w_1(0) + t_2 w_2(0) + g_{12,3} v_3(0) &= 0, \nonumber \\
    -g_{12,3} v_{12}(0) + t_3 w_3(0) &= 0. 
\end{align}
In a manner similar to other junctions, we can count Majorana junction modes.
The result is summarized in Table \ref{table_edgemode}.

\section{Tri-Junctions of 2D spin lattices} \label{section_2djunction}
In this section, we construct exactly solvable tri-junctions of two-dimensional (2D) spin lattices and examine their properties.

\subsection{2D spin lattice} \label{subsection_2D}
\begin{figure}
    \centering
    \includegraphics[width=5cm]{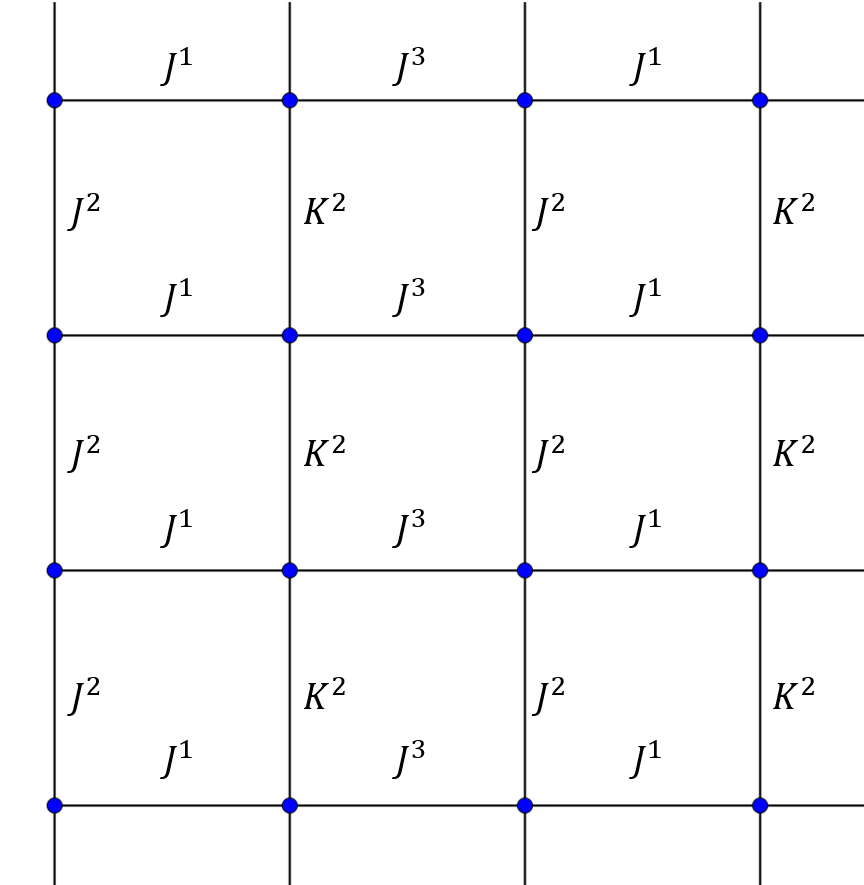}
    \caption{Half-plane square lattice.
    A $SO(5)$-spin is placed on each vertex.
    Labels on edges represent the exchange constants of spins.}
    \label{fig:2D_hopping}
\end{figure}
\begin{figure*}[tbp]
    \begin{minipage}{0.48\linewidth}
        \centering
        \includegraphics[width=85mm]{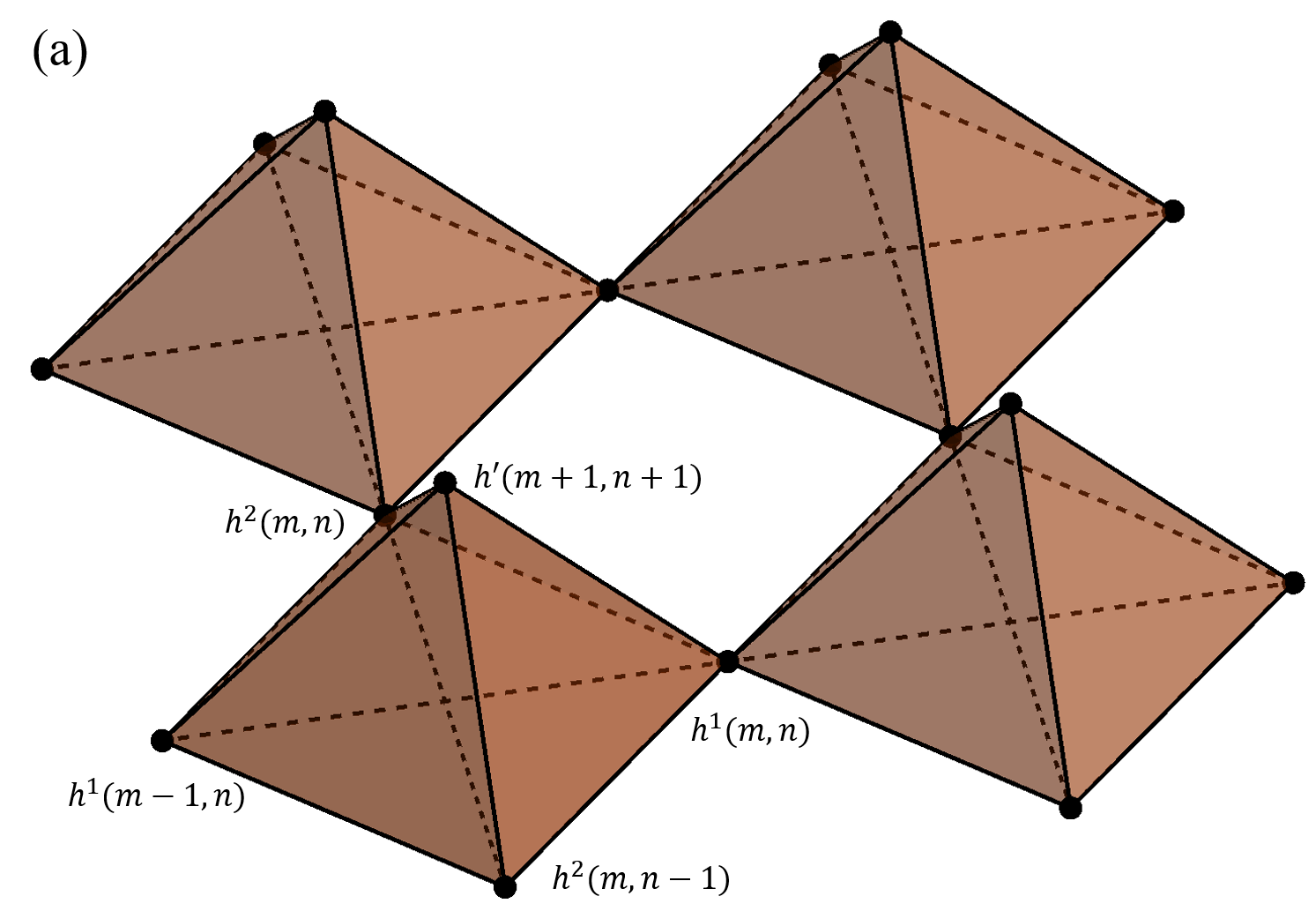}
    \end{minipage} 
    \begin{minipage}{0.48\linewidth}
        \centering
        \includegraphics[width=85mm]{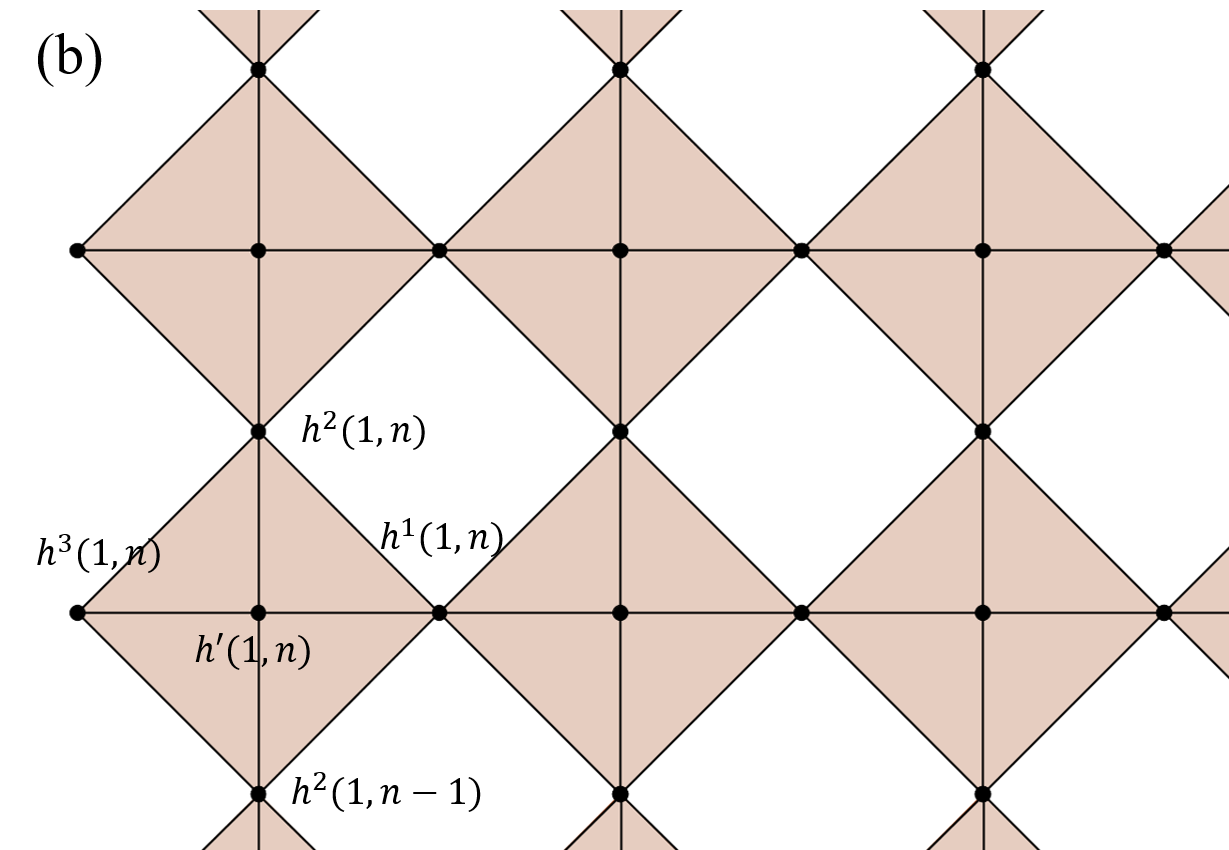}
    \end{minipage} 
    \caption{The CG of the $SO(5)$-spin square lattice.
    The right figure (b) is a simplified version of the left one (a), which we use in Fig.\ref{figs:2D_junctions}.}
    \label{figs:CG_plane}
\end{figure*}

As a 2D spin lattice system,
we consider an $SO(5)$-spin model on the square lattice.  
We put the $SO(5)$ Dirac matrices
\begin{align}
    \alpha^a(m,n)&=\sigma^a(m,n) \otimes \tau^x(m,n) \quad (a=1,2,3), \nonumber \\
    \alpha^4(m,n)&=\sigma^0(m,n) \otimes \tau^z(m,n), \nonumber \\
    \alpha^5(m,n)&=\sigma^0(m,n) \otimes \tau^y(m,n),
\end{align}
on each vertex of the half-plane square lattice in Fig. \ref{fig:2D_hopping}, 
and introduce the Hamiltonian as
\begin{align}
    H=-\sum_{m=1}^\infty \sum_{n\in \integers}[
    &J^1 h^1(2m-1,n) +J^3 h^1(2m,n) \nonumber \\
    &+J^2 h^2(2m-1,n) +K^2 h^2(2m,n) \nonumber \\
    &+J^5 h'(m,n)
    ], \nonumber \\
    -t \sum_{n \in \integers}& h^3(1,n) ,
\end{align}
where
\begin{align}
    h^1(2m-1,n) &= \alpha^1(2m-1,n) \alpha^1(2m,n), \nonumber \\
    h^1(2m,n) &= \alpha^3(2m,n) \alpha^3(2m+1,n), \nonumber \\
    h^2(2m-1,n) &= \alpha^2(2m-1,n) \alpha^4(2m-1,n+1), \nonumber \\
    h^2(2m,n) &= \alpha^2(2m,n) \alpha^4(2m,n+1), \nonumber \\
    h'(m,n) &= \alpha^5(m,n), \quad h^3(1,n) = \alpha^3(1,n).
\end{align}
The CG of this Hamiltonian is given by Fig. \ref{figs:CG_plane}.
Regarding each square pyramid and each endpoint in Fig. \ref{figs:CG_plane} as a 4-simplex and a 0-simplex, respectively, we can identify the CG as an SPSC.
Thus, each term of the Hamiltonian is equivalent to an MQF:
\begin{align} \label{2D_transformed}
    h^1(m,n)
    &=-i \epsilon^1(m,n) \varphi(m,n) \varphi(m+1,n), \nonumber \\
    h^2(m,n)
    &=-i \epsilon^2(m,n) \varphi(m,n) \varphi(m+1,n), \nonumber \\
    h'(m,n)&=
    -i \epsilon'(m,n) \varphi(m,n) \varphi'(m,n), \nonumber \\
    h^3(1,n)&=
    -i \epsilon^3(1,n) \varphi(0,n) \varphi(1,n),
\end{align}
where $\epsilon^d(m,n)$'s and $\epsilon'(m,n)$'s are sign factors.
Using the gauge transformation, we can set
\begin{align}
    \epsilon^1(m,n) 
    = \epsilon^2(1,n)
    = \epsilon^3(1,n)
    = \epsilon'(m,n)
    = 1
\end{align}
for any $m=,1,2,\ldots$ and $n \in \integers$.
The remaining factors are given by the conserved quantities on the holes in Fig. \ref{figs:CG_plane}.
Namely, we have the conserved quantity $c(m,n)$
\begin{align}
    c(m,n):=&h^2(m,n) h^1(m,n+1) h^2(m+1,n) h^1(m,n) \nonumber \\
    =&\epsilon^2(m,n) \epsilon^2(m+1,n), 
\end{align}
which determine the remaining factors.
From the Lieb's theorem \cite{lieb2004flux,macris1996flux}, we have $c(m,n)=-1$ for the ground state of this model, 
which fixes the remaining factors as
\begin{align}
    \epsilon^2(m,n)=(-1)^{m+1}.
\end{align}
Hereafter we set $J^5=0$ for simplicity.

Since our model has lattice translation symmetry in the $n$-direction, it is convenient to perform the Fourier transformation in the $n$-direction:
\begin{align}
    \varphi(m,n)=
    \frac{1}{\sqrt{\pi}} \int_{-\pi}^{\pi} dk_y a(m,k_y)e^{-ik_yn} \quad (m=0,1,2,\ldots),
\end{align}
where $k_y$ is the momentum in the $n$-direction, and $a(m,k_y)$ satisfies
\begin{align}
    a(m,k_y)^\dagger &= a(m,-k_y), \nonumber \\
    \{ a(m,k_y), a(m',k_y')^\dagger \}
    &=\delta_{m,m'} \delta(k_y-k_y').
\end{align}
After the Fourier transformation, we get
\begin{align}
    H= \int_{-\pi}^\pi dk_y \Vec{a}(k_y)^\dagger \mathcal{H}(k_y) \Vec{a}(k_y),
\end{align}
where
\begin{align}
    \Vec{a}(k_y)=
    \begin{pmatrix}
        a(0,k_y) & a(1,k_y) & a(2,k_y) & \cdots
    \end{pmatrix}^T
\end{align}
and
\begin{align} \label{2D_matrix}
    &\mathcal{H}(k_y)= \nonumber \\
    &\begin{pmatrix}
        0 & it \\
        -it & 2J^2\sin k_y & iJ^1 &\\
         & -iJ^1 & -2K^2\sin k_y & iJ^3 &\\
         &  & -iJ^3 & 2J^2\sin k_y & iJ^1 \\
         &  &  & -iJ^1 & \ddots & \ddots \\
         & & & & \ddots
    \end{pmatrix}.
\end{align}

We solve the energy eigenequation of $\mathcal{H}(k_y)$.
Let $E(k_y)$ be an eigenenergy of $\mathcal{H}(k_y)$ with
an eigenvector
\begin{align}
    \Vec{v}(k_y)=
    \begin{pmatrix}
        v(0,k_y) & w(0,k_y) & v(1,k_y) & w(1,k_y) & \cdots 
    \end{pmatrix}^T.
\end{align}
Then, the eigenequation gives
\begin{align}
    &itw(0,k_y) =E v(0,k_y) \label{2d_eigen_1}, \\
    &-itv(0,k_y) +2J^2\sin k_y  w(0,k_y) \nonumber \\
    & \quad +iJ^1v(1,k_y) = Ew(0,k_y), \label{2d_eigen_2}
    \\
    &-iJ^1w(n-1,k_y) -2K^2 \sin k_y v(n,k_y) \nonumber \\
    & \quad +iJ^3 w(n,k_y) = Ev(n,k_y), \label{2d_eigen_3} 
    \\
    &-iJ^3 v(n,k_y) +2J^2\sin k_y  w(n,k_y) \nonumber \\
    & \quad + iJ^1v(n+1,k_y) = Ew(n,k_y), \label{2d_eigen_4}
\end{align}
with $n=1,2,\ldots$.
To solve the eigenequation, we introduce the generating functions,
\begin{align}
    V(z)=\sum_{n=1}^\infty z^n v(n,k_y), \quad
    W(z)=\sum_{n=1}^\infty z^n w(n,k_y).
\end{align}
Then, from Eqs. (\ref{2d_eigen_2}), (\ref{2d_eigen_3}) and (\ref{2d_eigen_4}), we get
\begin{align}
    &\begin{pmatrix}
        E+2K^2\sin k_y & i(J^1z-J^3) \\
        i(J^3-iJ^1z^{-1} & E -2J^2\sin k_y
    \end{pmatrix}
    \begin{pmatrix}
        V \\
        W
    \end{pmatrix} \nonumber \\
    &=-w(0)
    \begin{pmatrix}
        iJ^1z \\
        E-2J^2\sin k_y
    \end{pmatrix}
    -v(0)
    \begin{pmatrix}
        0 \\
        it
    \end{pmatrix}.
\end{align}
Hence we obtain
\begin{align}
    V(z)&=
    \frac{-z}{J^1J^3 (z^2-2\kappa z +1)} \nonumber \\
    &\times \left[ t(J^1z-J^3)v(0,k_y) \right. \nonumber \\
    &\left. \quad + iJ^3(E-2J^2\sin k_y)w(0,k_y) \right], \nonumber \\
    W(z)&=
    \frac{-z}{J^1 J^3 (z^2-2\kappa z +1)} \nonumber \\
    &\times \left[ it(E+2K^2\sin k_y)v(0,k_y) \right. \nonumber \\ 
    &\left. \quad + (J^3-2J^1\kappa +J^1z)w(0,k_y) \right],
\end{align}
where
\begin{align}
    \kappa:=
    \frac{1}{2J^1J^3}[
    &(J^1)^2+(J^3)^2 \nonumber \\
    &-(E-2J^2\sin k_y)(E+2K^2\sin k_y)].
\end{align}

We can obtain the bulk spectrum as follows.
Let $z_\pm$ be the roots of $z^2 - 2\kappa z + 1$.
Then, in a manner similar to the transverse field Ising model, the bulk spectrum requires  $|z_\pm|=1$, i.e.  $z_\pm=e^{\pm ik_x}$ ($k_y\in (-\pi,\pi]$), which leads to
\begin{align}
    E
    =
    (J^2-K^2)\sin k_y \pm &\left[\{(J^2+K^2)\sin k_y \}^2 \right. \nonumber \\ 
    & \left. +(J^1)^2 + (J^3)^2 -2J^1J^3 \sin k_x \right].
\end{align}

\begin{figure*}[tbp]
    \begin{minipage}{0.48\linewidth}
        \centering
        \includegraphics[width=85mm]{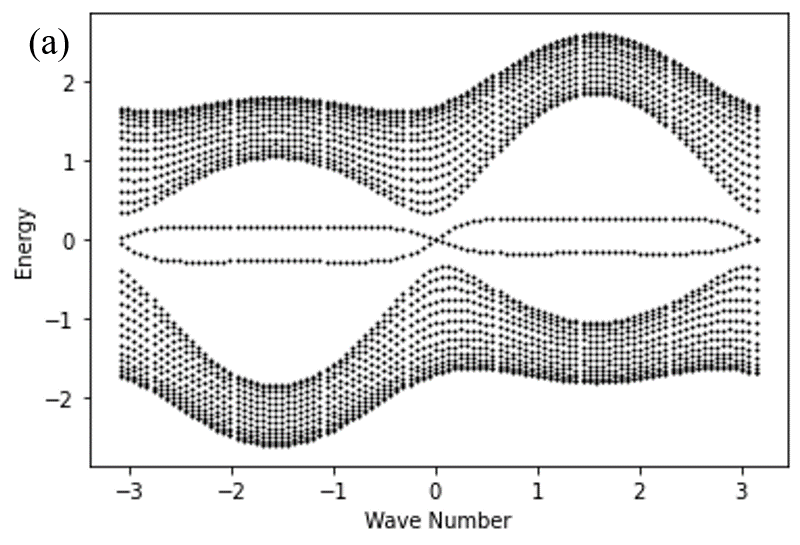}
    \end{minipage}
    \begin{minipage}{0.48\linewidth}
        \centering
        \includegraphics[width=85mm]{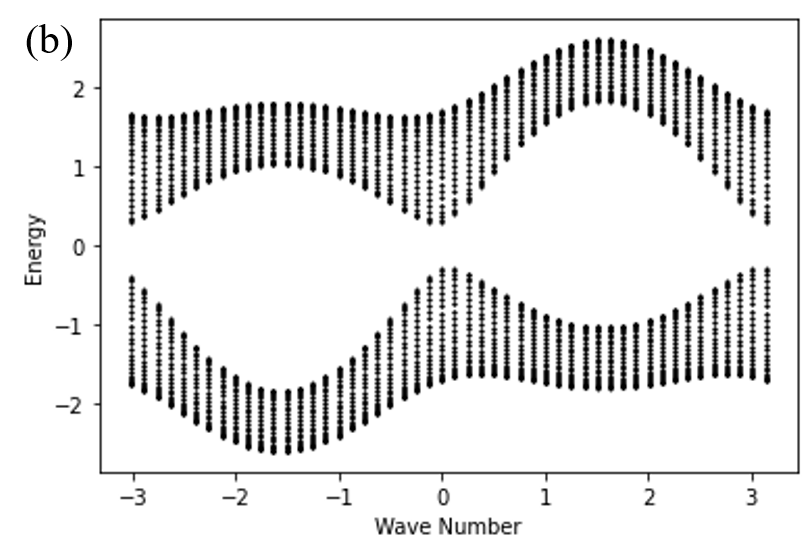}
    \end{minipage}
    \caption{The energy spectra of the $SO(5)$-spin lattice under (a) the open boundary condition (OBC) and (b) the periodic boundary condition (PBC) in the $m$-direction. The horizontal axis is the wave number $k_y$ in the $n$-direction. Here we take the model parameters as $J^1=1, J^2=0.9, J^3=0.7, K^2=0.5$, and $t=0.7$.
    Majorana modes along the tri-junction appear under the OBC.}
    \label{figs:2D_calculation}
\end{figure*}

For boundary zero modes with $E=0$,
the generating functions become
\begin{align} \label{2D_mother_calculated}
    V(z)=&
    \frac{-z}{J^1 J^3 (z^2-2\kappa z +1)} \nonumber \\
    &\times \left[ t(J^1z-J^3)v(0) - 2iJ^3J^2\sin k_y w(0,k_y) \right], \nonumber \\
    W(z)=&
    \frac{-z}{J^1 J^3 (z^2-2\kappa z +1)} \nonumber \\
    &\times \left[ 2itK^2\sin k_y  v(0) \right. \nonumber \\
    &\left. \quad+ (J^3-2J^1\kappa +J^1z)w(0,k_y) \right].
\end{align}
The terms involving $\sin k_y$ in Eq. (\ref{2D_mother_calculated}) requires $k_y=0,\pi$ to satisfy $|z_\pm| \neq 1$.
Under this condition, we get
\begin{align}
    V(z)
    &=\frac{tz}{J^1-J^3z}v(0,k_0)
    =\frac{t}{J^1}v(0) \sum_{n=1}^\infty \left( \frac{J^3z}{J^1} \right)^n, \nonumber \\
    W(z)
    &=\frac{J^1z}{J^3-J^1z}w(0,k_0)
    =w(0)\sum_{n=1}^\infty \left( \frac{J^1z}{J^3} \right)^n,
\end{align}
with $k_0=0,\pi$.
Moreover, Eq. (\ref{2d_eigen_1}) leads to $w(0,k_0)=0$, which means $W(z)=0$.
Then, from the $L^2$ condition for $v(n,k_0)$ of $V(z)$, we find that a zero energy edge state exists at $k_y=0,\pi$ when $J^1>J^3$.
In Fig. \ref{figs:2D_calculation} (a), we show the spectrum of the system under the open boundary condition in the $m$-direction,
which confirms the existence of the edge states for $J^1>J^3$.

Note that if we replace $J^1$ and $J^3$ with $J$ and $h$ respectively and set $k_y=0,\pi$ (or $J^2=K^2=0$), $\mathcal{H}$ coincides with the Hamiltonian of the transverse field Ising model in Eq. (\ref{TFIM_matrix}).

\subsection{Exactly Solvable Tri-Junctions of 2D Spins}
This section gives a class of exactly solvable tri-junctions of a 2D spin system by generalizing the arguments in Sec. \ref{sec:stj1d}. 
The model we consider is
\begin{align}
    H=H^{\mathrm{junc}}+\sum_{\mu=1}^3 H_\mu^{\mathrm{bulk}}.
\end{align}
Here $H^\bulk_\mu$ is the 2D $SO(5)$-spin lattice Hamiltonian, which is given by 
\begin{align}
    H_\mu^\bulk=-\sum_{m=1}^\infty \sum_{n\in \integers}[
    &J_\mu^1 h^1_\mu(2m-1,n) +J_\mu^3 h_\mu^1(2m,n) \nonumber \\
    &+J_\mu^2 h_\mu^2(2m-1,n) +K_\mu^2 h_\mu^2(2m,n)],
\end{align}
where
\begin{align}
    h_\mu^1(2m-1,n) &= \alpha_\mu^1(2m-1,n) \alpha_\mu^1(2m,n), \nonumber \\
    h_\mu^1(2m,n) &= \alpha_\mu^3(2m,n) \alpha_\mu^3(2m+1,n), \nonumber \\
    h_\mu^2(2m-1,n) &= \alpha_\mu^2(2m-1,n) \alpha_\mu^4(2m-1,n+1), \nonumber \\
    h_\mu^2(2m,n) &= \alpha_\mu^2(2m,n) \alpha_\mu^4(2m,n+1),
\end{align}
and the $SO(5)$ Dirac matrices are defined by
\begin{align}
    \alpha_\mu^a(m,n)&=\sigma_\mu^a(m,n) \otimes \tau_\mu^x(m,n), \quad (a=1,2,3)\nonumber \\
    \alpha_\mu^4(m,n)&=\sigma_\mu^0(m,n) \otimes \tau_\mu^z(m,n), \nonumber \\
    \alpha_\mu^5(m,n)&=\sigma_\mu^0(m,n) \otimes \tau_\mu^y(m,n).
\end{align}
$H^\junc$ is the junction Hamiltonian determined below.

Like tri-junctions of spin chains, we determine $H^\junc$ as follows.
Let $\Omega$ be a set whose elements $w$ are subsets of $\{1,2,3\}$ and satisfy $\bigsqcup_{\omega \in \Omega} \omega = \{1,2,3\}$. (See Eq. (\ref{omega_patterns}).)
Then, we consider the junction in the form of
\begin{align}
    H^\junc=
    &-t \sum_{\mu=1}^3 \sum_{n \in \integers} h_\mu^3(1,n) \nonumber \\
    &-\sum_{n \in \integers} \sum_{\omega,\omega' \in \Omega, \omega \neq \omega'} g_{\omega,\omega'} h_{\omega,\omega'}(n),
\label{eq:2djunc}
\end{align}
where the first term is the coupling to the 2D lattices, and the second is the intra-junction coupling that commutes with $H^\bulk$. 
We require that $h_\mu^3(1,n)$ and $h_{\omega,\omega'}(n)$ satisfy the same anti-commuatation relation as those of $h_\mu(1)$ and $h_{\omega,\omega'}(0)$ in Eq. (\ref{eq:junc}) for each $n$, and all the other relations are commutative. 

\begin{figure*}[tbp]
    \begin{minipage}{0.32\linewidth}
        \centering
        \includegraphics[width=5.5cm]{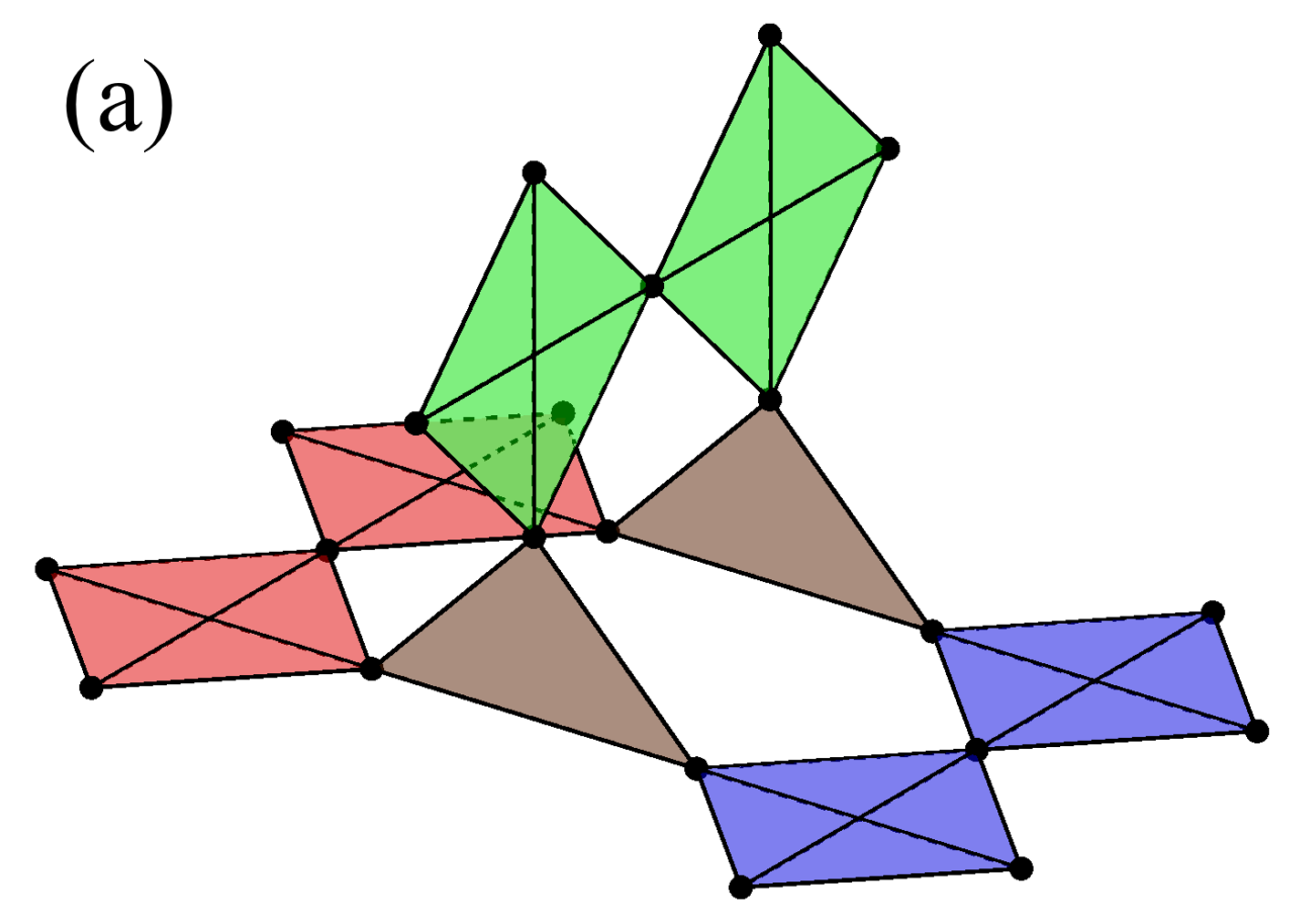}
    \end{minipage} 
    \begin{minipage}{0.32\linewidth}
        \centering
        \includegraphics[width=5.5cm]{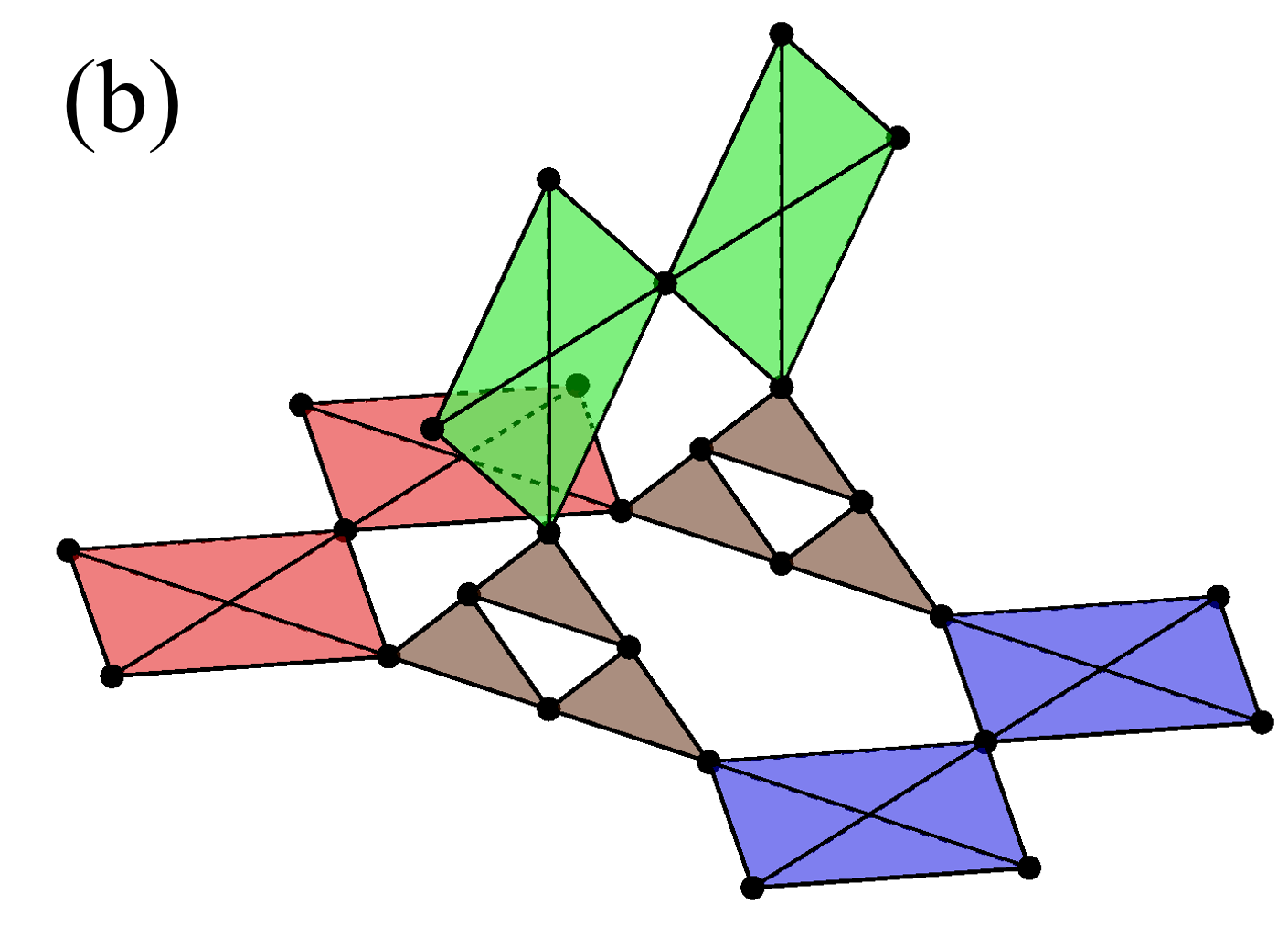}
    \end{minipage} 
    \begin{minipage}{0.32\linewidth}
        \centering
        \includegraphics[width=5.5cm]{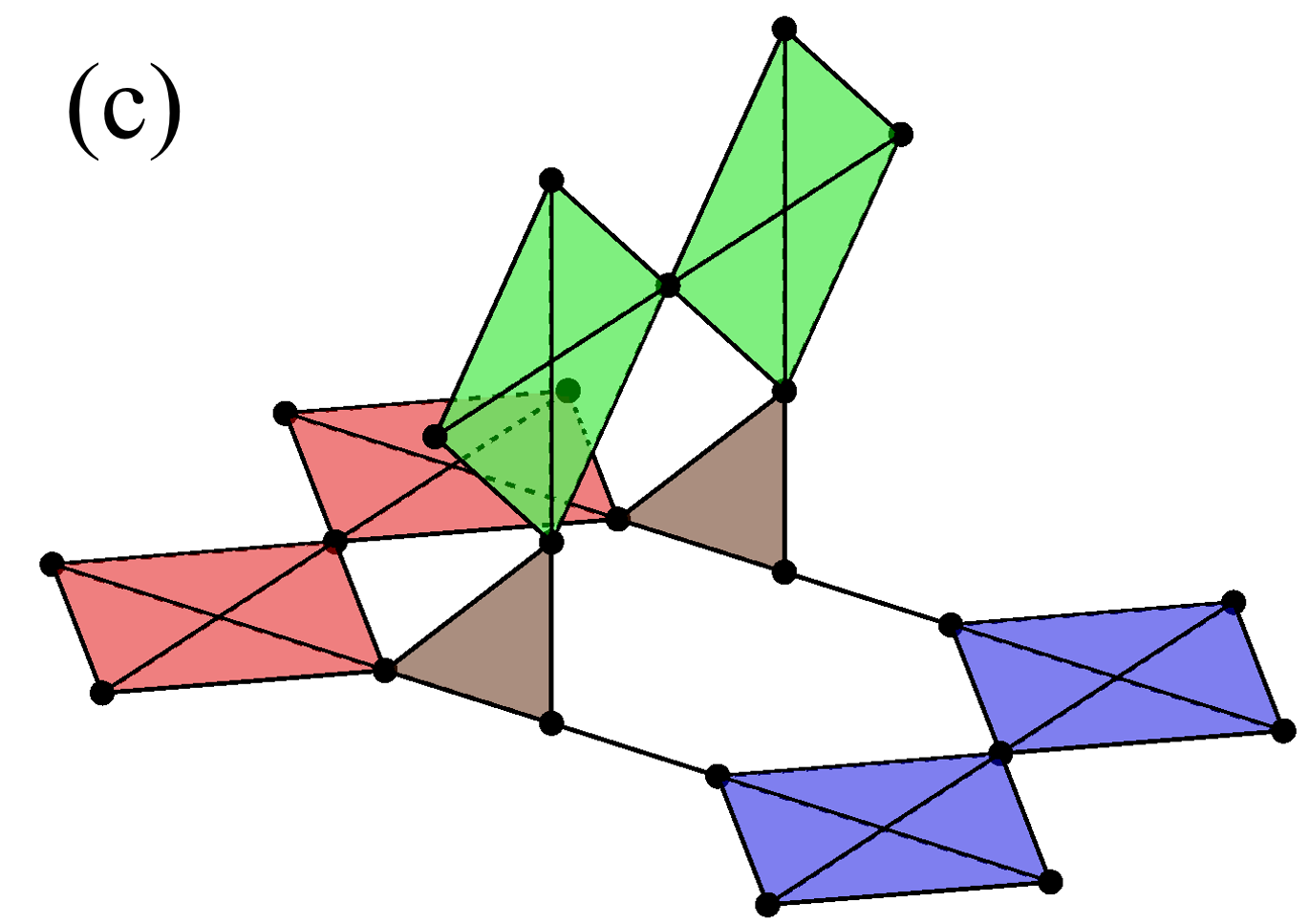}
    \end{minipage}
    \caption{Three patterns of tri-junctions of 2D $SO(5)$-spin lattices. (a) (3)-tri-junction. (b) (1+1+1)-tri-junction. (c) (2+1)-tri-junction.}
    \label{figs:2D_junctions}
\end{figure*}

The resultant junction is exactly solvable.
In Fig. \ref{figs:2D_junctions}, we show CGs of $H$ satisfying the condition above.
Corresponding to three different $\Omega$ in Eq. (\ref{omega_patterns}), 
we obtain three different patterns of CGs, all of which are SPSCs.

In this method, we have the same MQF of $H_\mu^\bulk$ as Eq. (\ref{2D_transformed}):
\begin{align} 
    h_\mu^1(m,n)
    &=-i \varphi_\mu(m,n) \varphi_\mu(m+1,n), \nonumber \\
    h_\mu^2(m,n)
    &=-i (-1)^{m+1} \varphi_\mu(m,n) \varphi_\mu(m+1,n), 
\end{align}
where we fix the sign factors using the Lieb's theorem. 
As a result, we have the same eigenequation for each of the 2D spin lattices,
\begin{align}
    &-iJ_\mu^1 w_\mu(n-1,k_y) -2K_\mu^2 \sin k_y  v_\mu(n,k_y) 
    \nonumber\\
    & \quad + iJ_\mu^3 w_\mu(n,k_y) = Ev_\mu(n,k_y), \nonumber\\
    &-iJ_\mu^3 v_\mu(n,k_y) +2J_\mu^2\sin k_y  w_\mu(n,k_y)
    \nonumber\\
    &\quad + iJ_\mu^1v_\mu(n+1,k_y)  = Ew_\mu(n,k_y),
    \label{eq:bulk2d}
\end{align}
with $n=1,2,\ldots$.
Thus we can solve the obtained junctions  by using the generating functions in a manner similar to Sec. \ref{subsection_2D}.

\subsection{Majorana junction states}

\begin{table*}[htbp]
    \centering
    \caption{
    The number of Majorana modes along the 2D $(3)$-, $(1+1+1)$-, and $(2+1)$-tri-junctions of the 2D $SO(5)$ spin lattices.
    }
    \begin{tabular}{c|ccc} \hline \hline
        Conditions & $(3)$-tri-junction & $(1+1+1)$-tri-junction & $(2+1)$-tri-junction  \\ \hline
        $J^3_1<J^1_1, J^3_2<J^1_2, J^3_3<J^1_3$ & 1 & 1 & 0 \\
        $J^3_1>J^1_1, J^3_2<J^1_2, J^3_3<J^1_3$ & 0 & 0 & 1 \\
        $J^3_1<J^1_1, J^3_2<J^1_2, J^3_3>J^1_3$ & 0 & 0 & 1 \\
        $J^3_1>J^1_1, J^3_2>J^1_2, J^3_3<J^1_3$ & 1 & 1 & 2 \\
        $J^3_1>J^1_1, J^3_2<J^1_2, J^3_3>J^1_3$ & 1 & 1 & 0 \\
        $J^3_1>J^1_1, J^3_2>J^1_2, J^3_3>J^1_3$ & 2 & 0 & 1 \\ \hline \hline
    \end{tabular}
    \label{table_edgemode_2D}
\end{table*}

\subsubsection{$(3)$-Tri-Junction}
First, we consider the $(3)$-tri-junction in Fig. \ref{figs:2D_junctions}(a) i.e., the tri-junction with $\Omega=\{123\}$.
To realize this junction, we consider
\begin{align}
    h_\mu^3(1,n)=\alpha_\mu^3(1,n)\zeta_{\mu+1}^3(1,n),
 \quad 
  h_{\omega,\omega'}(n)=0
\end{align}
in Eq.(\ref{eq:2djunc}), 
where 
\begin{align}
\zeta_\mu^3(1,n)=-\sigma_\mu^3(1,n)\otimes \tau_\mu^z(1,n)   
\end{align}
with $\zeta_{4}^3(1,n)=\zeta_1^3(1,n)$.
In an MQF, we have
\begin{align}
    h_\mu^3(1,n)=-i \epsilon_\mu(n) \varphi_{123}(0,n) \varphi_\mu(1,n),
\end{align}
where $\epsilon_\mu(n)$ is a sign factor, which we set
$\epsilon_\mu(n)=1$ using the Leib's theorem.
After the Fourier transformation, the junction Hamiltonian is recast into
\begin{align}
    H^\junc=
    \sum_{\mu=1}^3 it_\mu \int_{-\pi}^\pi dk_y a_{123}(0,k_y)^\dagger a_\mu(1,k_y) + \mathrm{h.c.}.
\end{align}
Thus, the eigenequation for the junction part is
\begin{align}
    &\sum_{\mu=1}^3 it_\mu w_\mu(0,k_y) = E v_{123}(0,k_y), \nonumber \\
    &-it_\mu v_{123}(0,k_y) + 2 J_\mu^2 \sin k_y  w_\mu(0,k_y) + iJ_\mu^1 v_\mu(1,k_y) 
    \nonumber\\
    & \quad = Ew_\mu(0,k_y),
\label{eq:(3)2d}
\end{align}
with $\mu=1,2,3$.
From this equation and Eq. (\ref{eq:bulk2d}), 
the generating functions
\begin{align}
    V_\mu(z) = \sum_{n=1}^\infty v_\mu(n,k_y)z^n, \quad
    W_\mu(z) = \sum_{n=1}^\infty w_\mu(n,k_y)z^n
\end{align}
are calculated as
\begin{align}
    V_\mu(z)=
    &\frac{-z}{J_\mu^3 (z^2-2\kappa_\mu z +1)} \nonumber \\
    &\times \left[ t_\mu(J_\mu^1z-J_\mu^3)v_\mu(0,k_y) \right. \nonumber \\
    & \left.\quad + iJ_\mu^3(E-2J_\mu^2\sin k_y)w(0,k_y) \right], \nonumber \\
    W_\mu(z)=
    &\frac{-z}{J_\mu^1 (z^2-2\kappa_\mu z +1)} \nonumber \\
    &\times \left[ it_\mu(E+2K_\mu^2\sin k_y)v(0,k_y) \nonumber \right. \\
    & \left. \quad + (J_\mu^3-2J_\mu^1\kappa_\mu +J_\mu^1z)w(0,k_y) \right],
\end{align}
where
\begin{align}
    \kappa_\mu:=
    \frac{1}{2J_\mu^1 J_\mu^3} &\left[(J_\mu^1)^2+(J_\mu^3)^2 \nonumber \right. \\
    & \left. -(E-2J_\mu^2 \sin k_y)(E+2K_\mu^2\sin k_y )\right].
\end{align}

To consider Majorana junction modes, we set $E=0$ and $k_y=0,\pi$.
Under this condition, the generating functions read
\begin{align}
    V_\mu(z)&=\frac{t_\mu}{J_\mu^1}v_{123}(0,k_0)\sum_{n=1}^\infty \left( \frac{J_\mu^3z}{J_\mu^1} \right)^n, \nonumber \\
    W_\mu(z)&=w_\mu(0,k_0)\sum_{n=1}^\infty \left( \frac{J_\mu^1 z}{J_\mu^3} \right)^n \quad (k_0=0,\pi),
\end{align}
which coincide with $V_\mu(z)$ and $W_{\mu}(z)$ in Eq.(\ref{1djunction_edge_(3)_mother}) with $J_\mu=J_\mu^1$ and $h_\mu=J_\mu^3$. Moreover, for $E=0$ and $k_y=0,\pi$, Eq. (\ref{eq:(3)2d}) leads to the same constraint as Eq. (\ref{1djunction_edge_(3)_relation}). 
Thus, we can count the number of Majorna junction modes similarly. We summarize the result in Table \ref{table_edgemode_2D}.

\subsubsection{$(1+1+1)$-Tri-Junction}
An example of the $(1+1+1)$-tri-junction in Fig. \ref{figs:2D_junctions} (b), i.e., the tri-junction with $\Omega = \{ \{1\}, \{2\}, \{3\} \}$,
we consider
\begin{align}
    h_\mu^3(1,n)=\alpha^3_\mu(1,n), \quad
    h_{\mu,\nu}(n) = \zeta_\mu^3(1,n) S^\rho(n) \zeta_\nu^3(1,n),
\end{align}
where we take $\omega=\mu$ and $\omega'=\nu$ in $h_{\omega,\omega'}(b)$ and $S^\rho(n)$ is an additional spin.
We find that the CG of this junction gives Fig. \ref{figs:2D_junctions}(b), and thus they are transformed into
\begin{align}
    h_\mu^3(1,n)&= -i \varphi_\mu(0,n) \varphi_\mu(1,n), \nonumber \\ 
    h_{\mu,\nu}(n)&= -i \epsilon_{\mu\nu}(n) \varphi_\mu(0,n) \varphi_\nu(0,n)
\end{align}
where we fix the sign factor in $h_\mu^3(1,n)$ using the gauge transformation.

In the present case, any choice of the remaining sign factor $\epsilon_{\mu\nu}$ does not satisfy the Leib's theorem at holes of the junction.
Thus, we just assume that $\epsilon_{12}(n) = \epsilon_{23}(n) = \epsilon_{31}(n) = 1$.
After the Fourier transformation, the junction is represented as
\begin{align}
    H^\junc=i\int_{-\pi}^\pi dk_y &\left[
    \sum_{\mu=1}^3 t_\mu a_\mu(0,k_y)^\dagger a_\mu(1,k_y) \right. \nonumber \\
    &+g_{12} a_1(0,k_y)^\dagger a_2(0,k_y) \nonumber \\
    &+g_{23} a_2(0,k_y)^\dagger a_3(0,k_y) \nonumber \\
    &\left.+g_{31} a_3(0,k_y)^\dagger a_1(0,k_y)
    \right],
    + \mathrm{h.c.}
\end{align}
which gives the eigenequation of the junction part as
\begin{align}
    &it_1 w_1(0,k_y) + ig_{12}v_2(0,k_y) -ig_{31} v_3(0,k_y) 
    \nonumber\\
    &=E v_1(0,k_y), \nonumber \\
    &-ig_{12}v_1(0,k_y) + it_2w_2(0,k_y) + ig_{23} v_3(0,k_y) 
    \nonumber\\
    &= E v_2(0,k_y), \nonumber \\
    &ig_{31} v_1(0,k_y) - ig_{23}v_2(0,k_y) + it_3 w_3(0,k_y) 
    \nonumber\\
    &= E v_3(0,k_y), \nonumber \\
    &-it_\mu v_\mu(0,k_y) +2J_\mu^2 \sin k_y w_\mu(0,k_y) +J_\mu^1 v_\mu(1,k_y) 
    \nonumber\\
    &= E w_\mu(0,k_y).
\end{align}
with $\mu=1,2,3$.
For $k_y=0,\pi$, the above equation coincides with Eq.(\ref{eq:(1+1+1)-1Djunc}) with $\lambda=-iE$, $J_\mu=J_\mu^1$ and $h_\mu=J_\mu^3$,
Therefore, Majorana junction modes appear in a manner similar to  
the  $(1+1+1)$-tri-junction in Sec.\ref{sec:(1+1+1)1D}.

Since Lieb's theorem can not apply to the present model, we have also examined the case with a different sign factor
\begin{align}
    \epsilon_{12}(n)
    =\epsilon_{13}(n)
    =\epsilon_{23}(n)
    =(-1)^n,
\end{align}
which gives the junction Hamiltonian as
\begin{align}
    H^\junc= \int_{-\pi}^\pi dk_y &\left[\sum_{\mu,\nu} 
    g_{\mu,\nu} a_\mu(0,k_y-\pi)^\dagger a_\nu(0,k_y) \right.
     \nonumber \\
    &\left. +\sum_{\mu} t_\mu a_\mu(0,k_y)^\dagger a_\mu(1,k_y)
    \right] + \mathrm{h.c.}.
\end{align}
Then, we find that Majorana junction modes appear in exactly the same manner as the above. 

We summarize the result in Table \ref{table_edgemode_2D}.

\subsubsection{$(2+1)$-Tri-Junction}
For the $(2+1)$-tri-junction in Fig.\ref{figs:2D_junctions} (c), that is, the tri-junction with $\Omega = \{ 
\{12\}, \{3\} \}$,
we consider the following terms in the junction
\begin{align}
    h_1^3(1,n) &= \alpha_1^3(1,n) \zeta_{2}^3(1,n), \quad
    h_2^3(1,n) = \alpha_2^3(1,n) \nonumber \\
    h_3^3(1,n) &= \alpha_3^3(1,n), \quad
    h_{12,3}(n)= \zeta_1^3(1,n) \zeta_2^3(1,n) \zeta_3^3(1,n),
\end{align}
which reproduce the CG in Fig. \ref{figs:2D_junctions} (c).
In MQFs, they are given by
\begin{align}
    h_1^3(1,n) &= -i \epsilon_{1}(n) \varphi_{12}(0,n) \varphi_1(1,n), \nonumber \\
    h_2^3(1,n) &= -i \epsilon_2(n) \varphi_{12}(0,n) \varphi_2(1,n) \nonumber \\
    h_3^3(1,n) &= -i \varphi_{3}(0,n) \varphi_3(1,n), \nonumber \\
    h_{12,3}(n) &= -i \varphi_{12}(0,n) \varphi_3(0,n),
\end{align}
where the sign factors in $h_3^3(1,n)$ and $h_{12,3}(n)$ are fixed by the gauge transformation.
The Lieb's theorem implies that 
\begin{align}
    \epsilon_1(n)=\epsilon_2(n)=(-1)^n,
\end{align}
from which we have the junction
\begin{align}
    H^\junc=
    \int_{-\pi}^\pi
    [
    &t_1a_{12}(0,k_y-\pi)^\dagger a_1(1,k_y) \nonumber \\
    &+t_2 a_{12}(0,k_y-\pi)^\dagger a_2(1,k_y) \nonumber \\
    &+t_2 a_3(0,k_y)^\dagger a_3(1,k_y) \nonumber \\
    &+g_{12,3} a_{12}(0,k_y)^\dagger a_3(0,k_y)
    ]
    + \mathrm{h.c.},
\end{align}
after the Fourier transformation with respect to $n$.
Then, repeating the same procedure as the other junction, 
we can examine how Majorna junction modes appear.
We summarize the result in Table \ref{table_edgemode_2D}.

\section{Discussion} \label{section_discussion}
This paper studies tri-junctions of one- and two-dimensional spin systems through exactly solvable models. We map the spin-tri-junctions into free Majorana fermion systems and argue the condition for the appearance of Majorana zero modes on the junctions. Our analysis reveals that details of local terms in the junctions crucially affect the presence or absence of Majorana zero modes. In particular, we find that Majorana zero modes may appear at the junction even when the bulk spin system does not support any boundary Majorana fermions.
For instance, the high magnetic field phase of the transverse field Ising spin chains hosts a Majorana zero mode when one considers the (3)-tri-junction (I) or the (2+1)-tri-junction. 
We also find that the tri-junctions of 2D spin systems exhibit the same phenomenon. We note that such a phenomenon never happens for the emergent Majorana fermions in topological phases. In the case of topological phases, the topological number ensures the presence and absence of Majorana fermions; thus, local terms do not affect them.

Whereas we consider the tri-junctions of the transverse field Ising spin chains and the 2D $SO(5)$-spin lattices, the generalization to other solvable spin systems, such as the XY model or the Kitaev spin model on the honeycomb lattice is straightforward. 
We can also generalize the argument to higher junctions, as discussed in Sec.\ref{sec:stj1d}. 
We hope to report on the results in these directions in the future.

\section*{Acknowledgements}
This  work  was  supported  by JST CREST Grant No. JPMJCR19T2 and KAKENHI Grant  No. JP20H00131.
M.O. was supported by a Grant-in-Aid for JSPS KAKENHI Grant No. JP22J15259.

\bibliography{references}

\begin{thebibliography}{43}%
\makeatletter
\providecommand \@ifxundefined [1]{%
 \@ifx{#1\undefined}
}%
\providecommand \@ifnum [1]{%
 \ifnum #1\expandafter \@firstoftwo
 \else \expandafter \@secondoftwo
 \fi
}%
\providecommand \@ifx [1]{%
 \ifx #1\expandafter \@firstoftwo
 \else \expandafter \@secondoftwo
 \fi
}%
\providecommand \natexlab [1]{#1}%
\providecommand \enquote  [1]{``#1''}%
\providecommand \bibnamefont  [1]{#1}%
\providecommand \bibfnamefont [1]{#1}%
\providecommand \citenamefont [1]{#1}%
\providecommand \href@noop [0]{\@secondoftwo}%
\providecommand \href [0]{\begingroup \@sanitize@url \@href}%
\providecommand \@href[1]{\@@startlink{#1}\@@href}%
\providecommand \@@href[1]{\endgroup#1\@@endlink}%
\providecommand \@sanitize@url [0]{\catcode `\\12\catcode `\$12\catcode
  `\&12\catcode `\#12\catcode `\^12\catcode `\_12\catcode `\%12\relax}%
\providecommand \@@startlink[1]{}%
\providecommand \@@endlink[0]{}%
\providecommand \url  [0]{\begingroup\@sanitize@url \@url }%
\providecommand \@url [1]{\endgroup\@href {#1}{\urlprefix }}%
\providecommand \urlprefix  [0]{URL }%
\providecommand \Eprint [0]{\href }%
\providecommand \doibase [0]{http://dx.doi.org/}%
\providecommand \selectlanguage [0]{\@gobble}%
\providecommand \bibinfo  [0]{\@secondoftwo}%
\providecommand \bibfield  [0]{\@secondoftwo}%
\providecommand \translation [1]{[#1]}%
\providecommand \BibitemOpen [0]{}%
\providecommand \bibitemStop [0]{}%
\providecommand \bibitemNoStop [0]{.\EOS\space}%
\providecommand \EOS [0]{\spacefactor3000\relax}%
\providecommand \BibitemShut  [1]{\csname bibitem#1\endcsname}%
\let\auto@bib@innerbib\@empty
\bibitem [{\citenamefont {Jordan}\ and\ \citenamefont
  {Wigner}(1928)}]{jordan1928pauli}%
  \BibitemOpen
  \bibfield  {author} {\bibinfo {author} {\bibfnamefont {Pascual}\ \bibnamefont
  {Jordan}}\ and\ \bibinfo {author} {\bibfnamefont {Eugene}\ \bibnamefont
  {Wigner}},\ }\bibfield  {title} {\enquote {\bibinfo {title} {Pauli’s
  equivalence prohibition},}\ }\href@noop {} {\bibfield  {journal} {\bibinfo
  {journal} {Z. Physik}\ }\textbf {\bibinfo {volume} {47}},\ \bibinfo {pages}
  {631} (\bibinfo {year} {1928})}\BibitemShut {NoStop}%
\bibitem [{\citenamefont {Onsager}(1944)}]{onsager1944crystal}%
  \BibitemOpen
  \bibfield  {author} {\bibinfo {author} {\bibfnamefont {Lars}\ \bibnamefont
  {Onsager}},\ }\bibfield  {title} {\enquote {\bibinfo {title} {Crystal
  statistics. i. a two-dimensional model with an order-disorder transition},}\
  }\href@noop {} {\bibfield  {journal} {\bibinfo  {journal} {Physical Review}\
  }\textbf {\bibinfo {volume} {65}},\ \bibinfo {pages} {117} (\bibinfo {year}
  {1944})}\BibitemShut {NoStop}%
\bibitem [{\citenamefont {Kaufman}(1949)}]{kaufman1949crystal}%
  \BibitemOpen
  \bibfield  {author} {\bibinfo {author} {\bibfnamefont {Bruria}\ \bibnamefont
  {Kaufman}},\ }\bibfield  {title} {\enquote {\bibinfo {title} {Crystal
  statistics. ii. partition function evaluated by spinor analysis},}\
  }\href@noop {} {\bibfield  {journal} {\bibinfo  {journal} {Physical Review}\
  }\textbf {\bibinfo {volume} {76}},\ \bibinfo {pages} {1232} (\bibinfo {year}
  {1949})}\BibitemShut {NoStop}%
\bibitem [{\citenamefont {Kaufman}\ and\ \citenamefont
  {Onsager}(1949)}]{kaufman1949crystal2}%
  \BibitemOpen
  \bibfield  {author} {\bibinfo {author} {\bibfnamefont {Bruria}\ \bibnamefont
  {Kaufman}}\ and\ \bibinfo {author} {\bibfnamefont {Lars}\ \bibnamefont
  {Onsager}},\ }\bibfield  {title} {\enquote {\bibinfo {title} {Crystal
  statistics. iii. short-range order in a binary ising lattice},}\ }\href@noop
  {} {\bibfield  {journal} {\bibinfo  {journal} {Physical Review}\ }\textbf
  {\bibinfo {volume} {76}},\ \bibinfo {pages} {1244} (\bibinfo {year}
  {1949})}\BibitemShut {NoStop}%
\bibitem [{\citenamefont {Nambu}(1995)}]{nambu1995note}%
  \BibitemOpen
  \bibfield  {author} {\bibinfo {author} {\bibfnamefont {Y{\^o}ichir{\^o}}\
  \bibnamefont {Nambu}},\ }\bibfield  {title} {\enquote {\bibinfo {title} {A
  note on the eigenvalue problem in crystal statistics},}\ }\href@noop {}
  {\bibfield  {journal} {\bibinfo  {journal} {Broken Symmetry: Selected Papers
  of Y Nambu}\ }\textbf {\bibinfo {volume} {13}},\ \bibinfo {pages} {1}
  (\bibinfo {year} {1995})}\BibitemShut {NoStop}%
\bibitem [{\citenamefont {Lieb}\ \emph {et~al.}(1961)\citenamefont {Lieb},
  \citenamefont {Schultz},\ and\ \citenamefont {Mattis}}]{lieb1961}%
  \BibitemOpen
  \bibfield  {author} {\bibinfo {author} {\bibfnamefont {Elliott}\ \bibnamefont
  {Lieb}}, \bibinfo {author} {\bibfnamefont {Theodore}\ \bibnamefont
  {Schultz}}, \ and\ \bibinfo {author} {\bibfnamefont {D.}~\bibnamefont
  {Mattis}},\ }\bibfield  {title} {\enquote {\bibinfo {title} {Two soluble
  models of an antiferromagnetic chain},}\ }\href {\doibase
  10.1016/0003-4916(61)90115-4} {\bibfield  {journal} {\bibinfo  {journal}
  {Annals of Physics}\ }\textbf {\bibinfo {volume} {16}},\ \bibinfo {pages}
  {407--466} (\bibinfo {year} {1961})}\BibitemShut {NoStop}%
\bibitem [{\citenamefont {Niemeijer}(1967)}]{niemeijer1967some}%
  \BibitemOpen
  \bibfield  {author} {\bibinfo {author} {\bibfnamefont {Th}~\bibnamefont
  {Niemeijer}},\ }\bibfield  {title} {\enquote {\bibinfo {title} {Some exact
  calculations on a chain of spins 12},}\ }\href@noop {} {\bibfield  {journal}
  {\bibinfo  {journal} {Physica}\ }\textbf {\bibinfo {volume} {36}},\ \bibinfo
  {pages} {377--419} (\bibinfo {year} {1967})}\BibitemShut {NoStop}%
\bibitem [{\citenamefont {Katsura}(1962)}]{katsura1962statistical}%
  \BibitemOpen
  \bibfield  {author} {\bibinfo {author} {\bibfnamefont {Shigetoshi}\
  \bibnamefont {Katsura}},\ }\bibfield  {title} {\enquote {\bibinfo {title}
  {Statistical mechanics of the anisotropic linear heisenberg model},}\
  }\href@noop {} {\bibfield  {journal} {\bibinfo  {journal} {Physical Review}\
  }\textbf {\bibinfo {volume} {127}},\ \bibinfo {pages} {1508} (\bibinfo {year}
  {1962})}\BibitemShut {NoStop}%
\bibitem [{\citenamefont {Pfeuty}(1970)}]{pfeuty1970one}%
  \BibitemOpen
  \bibfield  {author} {\bibinfo {author} {\bibfnamefont {Pierre}\ \bibnamefont
  {Pfeuty}},\ }\bibfield  {title} {\enquote {\bibinfo {title} {The
  one-dimensional ising model with a transverse field},}\ }\href@noop {}
  {\bibfield  {journal} {\bibinfo  {journal} {ANNALS of Physics}\ }\textbf
  {\bibinfo {volume} {57}},\ \bibinfo {pages} {79--90} (\bibinfo {year}
  {1970})}\BibitemShut {NoStop}%
\bibitem [{\citenamefont {Shankar}\ and\ \citenamefont
  {Murthy}(1987)}]{shankar1987nearest}%
  \BibitemOpen
  \bibfield  {author} {\bibinfo {author} {\bibfnamefont {R}~\bibnamefont
  {Shankar}}\ and\ \bibinfo {author} {\bibfnamefont {Ganpathy}\ \bibnamefont
  {Murthy}},\ }\bibfield  {title} {\enquote {\bibinfo {title} {Nearest-neighbor
  frustrated random-bond model in d= 2: Some exact results},}\ }\href@noop {}
  {\bibfield  {journal} {\bibinfo  {journal} {Physical Review B}\ }\textbf
  {\bibinfo {volume} {36}},\ \bibinfo {pages} {536} (\bibinfo {year}
  {1987})}\BibitemShut {NoStop}%
\bibitem [{\citenamefont {Minami}(2016)}]{minami2016solvable}%
  \BibitemOpen
  \bibfield  {author} {\bibinfo {author} {\bibfnamefont {Kazuhiko}\
  \bibnamefont {Minami}},\ }\bibfield  {title} {\enquote {\bibinfo {title}
  {Solvable hamiltonians and fermionization transformations obtained from
  operators satisfying specific commutation relations},}\ }\href@noop {}
  {\bibfield  {journal} {\bibinfo  {journal} {Journal of the Physical Society
  of Japan}\ }\textbf {\bibinfo {volume} {85}},\ \bibinfo {pages} {024003}
  (\bibinfo {year} {2016})}\BibitemShut {NoStop}%
\bibitem [{\citenamefont {Minami}(2017)}]{minami2017infinite}%
  \BibitemOpen
  \bibfield  {author} {\bibinfo {author} {\bibfnamefont {Kazuhiko}\
  \bibnamefont {Minami}},\ }\bibfield  {title} {\enquote {\bibinfo {title}
  {Infinite number of solvable generalizations of xy-chain, with cluster state,
  and with central charge c= m/2},}\ }\href@noop {} {\bibfield  {journal}
  {\bibinfo  {journal} {Nuclear Physics B}\ }\textbf {\bibinfo {volume}
  {925}},\ \bibinfo {pages} {144--160} (\bibinfo {year} {2017})}\BibitemShut
  {NoStop}%
\bibitem [{\citenamefont {Kitaev}(2006)}]{kitaev2006anyons}%
  \BibitemOpen
  \bibfield  {author} {\bibinfo {author} {\bibfnamefont {Alexei}\ \bibnamefont
  {Kitaev}},\ }\bibfield  {title} {\enquote {\bibinfo {title} {Anyons in an
  exactly solved model and beyond},}\ }\href@noop {} {\bibfield  {journal}
  {\bibinfo  {journal} {Annals of Physics}\ }\textbf {\bibinfo {volume}
  {321}},\ \bibinfo {pages} {2--111} (\bibinfo {year} {2006})}\BibitemShut
  {NoStop}%
\bibitem [{\citenamefont {Kitaev}\ and\ \citenamefont
  {Laumann}(2009)}]{kitaev2009topological}%
  \BibitemOpen
  \bibfield  {author} {\bibinfo {author} {\bibfnamefont {Alexei}\ \bibnamefont
  {Kitaev}}\ and\ \bibinfo {author} {\bibfnamefont {Chris}\ \bibnamefont
  {Laumann}},\ }\bibfield  {title} {\enquote {\bibinfo {title} {Topological
  phases and quantum computation},}\ }\href@noop {} {\bibfield  {journal}
  {\bibinfo  {journal} {Exact methods in low-dimensional statistical physics
  and quantum computing,” Lecture Notes of the Les Houches Summer School}\
  }\textbf {\bibinfo {volume} {89}},\ \bibinfo {pages} {101--125} (\bibinfo
  {year} {2009})}\BibitemShut {NoStop}%
\bibitem [{\citenamefont {Ryu}(2009)}]{ryu2009three}%
  \BibitemOpen
  \bibfield  {author} {\bibinfo {author} {\bibfnamefont {Shinsei}\ \bibnamefont
  {Ryu}},\ }\bibfield  {title} {\enquote {\bibinfo {title} {Three-dimensional
  topological phase on the diamond lattice},}\ }\href@noop {} {\bibfield
  {journal} {\bibinfo  {journal} {Physical Review B}\ }\textbf {\bibinfo
  {volume} {79}},\ \bibinfo {pages} {075124} (\bibinfo {year}
  {2009})}\BibitemShut {NoStop}%
\bibitem [{\citenamefont {Wu}\ \emph {et~al.}(2009)\citenamefont {Wu},
  \citenamefont {Arovas},\ and\ \citenamefont {Hung}}]{wu2009gamma}%
  \BibitemOpen
  \bibfield  {author} {\bibinfo {author} {\bibfnamefont {Congjun}\ \bibnamefont
  {Wu}}, \bibinfo {author} {\bibfnamefont {Daniel}\ \bibnamefont {Arovas}}, \
  and\ \bibinfo {author} {\bibfnamefont {Hsiang-Hsuan}\ \bibnamefont {Hung}},\
  }\bibfield  {title} {\enquote {\bibinfo {title} {$\gamma$-matrix
  generalization of the kitaev model},}\ }\href@noop {} {\bibfield  {journal}
  {\bibinfo  {journal} {Physical Review B}\ }\textbf {\bibinfo {volume} {79}},\
  \bibinfo {pages} {134427} (\bibinfo {year} {2009})}\BibitemShut {NoStop}%
\bibitem [{\citenamefont {Bochniak}\ and\ \citenamefont
  {Ruba}(2020)}]{bochniak2020bosonization}%
  \BibitemOpen
  \bibfield  {author} {\bibinfo {author} {\bibfnamefont {Arkadiusz}\
  \bibnamefont {Bochniak}}\ and\ \bibinfo {author} {\bibfnamefont
  {B{\l}a{\.z}ej}\ \bibnamefont {Ruba}},\ }\bibfield  {title} {\enquote
  {\bibinfo {title} {Bosonization based on clifford algebras and its gauge
  theoretic interpretation},}\ }\href@noop {} {\bibfield  {journal} {\bibinfo
  {journal} {Journal of High Energy Physics}\ }\textbf {\bibinfo {volume}
  {2020}},\ \bibinfo {pages} {1--36} (\bibinfo {year} {2020})}\BibitemShut
  {NoStop}%
\bibitem [{\citenamefont {Bochniak}\ \emph {et~al.}(2020)\citenamefont
  {Bochniak}, \citenamefont {Ruba}, \citenamefont {Wosiek},\ and\ \citenamefont
  {Wyrzykowski}}]{bochniak2020constraints}%
  \BibitemOpen
  \bibfield  {author} {\bibinfo {author} {\bibfnamefont {Arkadiusz}\
  \bibnamefont {Bochniak}}, \bibinfo {author} {\bibfnamefont {B{\l}a{\.z}ej}\
  \bibnamefont {Ruba}}, \bibinfo {author} {\bibfnamefont {Jacek}\ \bibnamefont
  {Wosiek}}, \ and\ \bibinfo {author} {\bibfnamefont {Adam}\ \bibnamefont
  {Wyrzykowski}},\ }\bibfield  {title} {\enquote {\bibinfo {title} {Constraints
  of kinematic bosonization in two and higher dimensions},}\ }\href@noop {}
  {\bibfield  {journal} {\bibinfo  {journal} {Physical Review D}\ }\textbf
  {\bibinfo {volume} {102}},\ \bibinfo {pages} {114502} (\bibinfo {year}
  {2020})}\BibitemShut {NoStop}%
\bibitem [{\citenamefont {Po}(2021)}]{po2021symmetric}%
  \BibitemOpen
  \bibfield  {author} {\bibinfo {author} {\bibfnamefont {Hoi~Chun}\
  \bibnamefont {Po}},\ }\bibfield  {title} {\enquote {\bibinfo {title}
  {Symmetric jordan-wigner transformation in higher dimensions},}\ }\href@noop
  {} {\bibfield  {journal} {\bibinfo  {journal} {arXiv preprint
  arXiv:2107.10842}\ } (\bibinfo {year} {2021})}\BibitemShut {NoStop}%
\bibitem [{\citenamefont {Li}\ and\ \citenamefont {Po}(2022)}]{li2022higher}%
  \BibitemOpen
  \bibfield  {author} {\bibinfo {author} {\bibfnamefont {Kangle}\ \bibnamefont
  {Li}}\ and\ \bibinfo {author} {\bibfnamefont {Hoi~Chun}\ \bibnamefont {Po}},\
  }\bibfield  {title} {\enquote {\bibinfo {title} {Higher-dimensional
  jordan-wigner transformation and auxiliary majorana fermions},}\ }\href@noop
  {} {\bibfield  {journal} {\bibinfo  {journal} {Physical Review B}\ }\textbf
  {\bibinfo {volume} {106}},\ \bibinfo {pages} {115109} (\bibinfo {year}
  {2022})}\BibitemShut {NoStop}%
\bibitem [{\citenamefont {Minami}(2019)}]{minami2019honeycomb}%
  \BibitemOpen
  \bibfield  {author} {\bibinfo {author} {\bibfnamefont {Kazuhiko}\
  \bibnamefont {Minami}},\ }\bibfield  {title} {\enquote {\bibinfo {title}
  {Honeycomb lattice kitaev model with wen--toric-code interactions, and anyon
  excitations},}\ }\href@noop {} {\bibfield  {journal} {\bibinfo  {journal}
  {Nuclear Physics B}\ }\textbf {\bibinfo {volume} {939}},\ \bibinfo {pages}
  {465--484} (\bibinfo {year} {2019})}\BibitemShut {NoStop}%
\bibitem [{\citenamefont {Nussinov}\ and\ \citenamefont
  {Ortiz}(2009)}]{nussinov2009bond}%
  \BibitemOpen
  \bibfield  {author} {\bibinfo {author} {\bibfnamefont {Zohar}\ \bibnamefont
  {Nussinov}}\ and\ \bibinfo {author} {\bibfnamefont {Gerardo}\ \bibnamefont
  {Ortiz}},\ }\bibfield  {title} {\enquote {\bibinfo {title} {Bond algebras and
  exact solvability of hamiltonians: Spin s= 1 2 multilayer systems},}\
  }\href@noop {} {\bibfield  {journal} {\bibinfo  {journal} {Physical Review
  B}\ }\textbf {\bibinfo {volume} {79}},\ \bibinfo {pages} {214440} (\bibinfo
  {year} {2009})}\BibitemShut {NoStop}%
\bibitem [{\citenamefont {Cobanera}\ \emph {et~al.}(2011)\citenamefont
  {Cobanera}, \citenamefont {Ortiz},\ and\ \citenamefont
  {Nussinov}}]{cobanera2011bond}%
  \BibitemOpen
  \bibfield  {author} {\bibinfo {author} {\bibfnamefont {Emilio}\ \bibnamefont
  {Cobanera}}, \bibinfo {author} {\bibfnamefont {Gerardo}\ \bibnamefont
  {Ortiz}}, \ and\ \bibinfo {author} {\bibfnamefont {Zohar}\ \bibnamefont
  {Nussinov}},\ }\bibfield  {title} {\enquote {\bibinfo {title} {The
  bond-algebraic approach to dualities},}\ }\href@noop {} {\bibfield  {journal}
  {\bibinfo  {journal} {Advances in physics}\ }\textbf {\bibinfo {volume}
  {60}},\ \bibinfo {pages} {679--798} (\bibinfo {year} {2011})}\BibitemShut
  {NoStop}%
\bibitem [{\citenamefont {Nussinov}\ \emph {et~al.}(2012)\citenamefont
  {Nussinov}, \citenamefont {Ortiz},\ and\ \citenamefont
  {Cobanera}}]{nussinov2012arbitrary}%
  \BibitemOpen
  \bibfield  {author} {\bibinfo {author} {\bibfnamefont {Zohar}\ \bibnamefont
  {Nussinov}}, \bibinfo {author} {\bibfnamefont {Gerardo}\ \bibnamefont
  {Ortiz}}, \ and\ \bibinfo {author} {\bibfnamefont {Emilio}\ \bibnamefont
  {Cobanera}},\ }\bibfield  {title} {\enquote {\bibinfo {title} {Arbitrary
  dimensional majorana dualities and architectures for topological matter},}\
  }\href@noop {} {\bibfield  {journal} {\bibinfo  {journal} {Physical Review
  B}\ }\textbf {\bibinfo {volume} {86}},\ \bibinfo {pages} {085415} (\bibinfo
  {year} {2012})}\BibitemShut {NoStop}%
\bibitem [{\citenamefont {Chen}\ and\ \citenamefont
  {Kapustin}(2019)}]{chen2019bosonization}%
  \BibitemOpen
  \bibfield  {author} {\bibinfo {author} {\bibfnamefont {Yu-An}\ \bibnamefont
  {Chen}}\ and\ \bibinfo {author} {\bibfnamefont {Anton}\ \bibnamefont
  {Kapustin}},\ }\bibfield  {title} {\enquote {\bibinfo {title} {Bosonization
  in three spatial dimensions and a 2-form gauge theory},}\ }\href@noop {}
  {\bibfield  {journal} {\bibinfo  {journal} {Physical Review B}\ }\textbf
  {\bibinfo {volume} {100}},\ \bibinfo {pages} {245127} (\bibinfo {year}
  {2019})}\BibitemShut {NoStop}%
\bibitem [{\citenamefont {Chen}(2020)}]{chen2020exact}%
  \BibitemOpen
  \bibfield  {author} {\bibinfo {author} {\bibfnamefont {Yu-An}\ \bibnamefont
  {Chen}},\ }\bibfield  {title} {\enquote {\bibinfo {title} {Exact bosonization
  in arbitrary dimensions},}\ }\href@noop {} {\bibfield  {journal} {\bibinfo
  {journal} {Physical Review Research}\ }\textbf {\bibinfo {volume} {2}},\
  \bibinfo {pages} {033527} (\bibinfo {year} {2020})}\BibitemShut {NoStop}%
\bibitem [{\citenamefont {Fendley}(2019)}]{fendley2019free}%
  \BibitemOpen
  \bibfield  {author} {\bibinfo {author} {\bibfnamefont {Paul}\ \bibnamefont
  {Fendley}},\ }\bibfield  {title} {\enquote {\bibinfo {title} {Free fermions
  in disguise},}\ }\href@noop {} {\bibfield  {journal} {\bibinfo  {journal}
  {Journal of Physics A: Mathematical and Theoretical}\ }\textbf {\bibinfo
  {volume} {52}},\ \bibinfo {pages} {335002} (\bibinfo {year}
  {2019})}\BibitemShut {NoStop}%
\bibitem [{\citenamefont {Chapman}\ and\ \citenamefont
  {Flammia}(2020)}]{chapman2020characterization}%
  \BibitemOpen
  \bibfield  {author} {\bibinfo {author} {\bibfnamefont {Adrian}\ \bibnamefont
  {Chapman}}\ and\ \bibinfo {author} {\bibfnamefont {Steven~T}\ \bibnamefont
  {Flammia}},\ }\bibfield  {title} {\enquote {\bibinfo {title}
  {Characterization of solvable spin models via graph invariants},}\
  }\href@noop {} {\bibfield  {journal} {\bibinfo  {journal} {arXiv preprint
  arXiv:2003.05465}\ } (\bibinfo {year} {2020})}\BibitemShut {NoStop}%
\bibitem [{\citenamefont {Prosko}\ \emph {et~al.}(2017)\citenamefont {Prosko},
  \citenamefont {Lee},\ and\ \citenamefont {Maciejko}}]{prosko2017simple}%
  \BibitemOpen
  \bibfield  {author} {\bibinfo {author} {\bibfnamefont {Christian}\
  \bibnamefont {Prosko}}, \bibinfo {author} {\bibfnamefont {Shu-Ping}\
  \bibnamefont {Lee}}, \ and\ \bibinfo {author} {\bibfnamefont {Joseph}\
  \bibnamefont {Maciejko}},\ }\bibfield  {title} {\enquote {\bibinfo {title}
  {Simple z 2 lattice gauge theories at finite fermion density},}\ }\href@noop
  {} {\bibfield  {journal} {\bibinfo  {journal} {Physical Review B}\ }\textbf
  {\bibinfo {volume} {96}},\ \bibinfo {pages} {205104} (\bibinfo {year}
  {2017})}\BibitemShut {NoStop}%
\bibitem [{\citenamefont {Yu}\ and\ \citenamefont
  {Wang}(2008)}]{yu2008exactly}%
  \BibitemOpen
  \bibfield  {author} {\bibinfo {author} {\bibfnamefont {Yue}\ \bibnamefont
  {Yu}}\ and\ \bibinfo {author} {\bibfnamefont {Ziqiang}\ \bibnamefont
  {Wang}},\ }\bibfield  {title} {\enquote {\bibinfo {title} {An exactly soluble
  model with tunable p-wave paired fermion ground states},}\ }\href@noop {}
  {\bibfield  {journal} {\bibinfo  {journal} {EPL (Europhysics Letters)}\
  }\textbf {\bibinfo {volume} {84}},\ \bibinfo {pages} {57002} (\bibinfo {year}
  {2008})}\BibitemShut {NoStop}%
\bibitem [{\citenamefont {Lee}\ \emph {et~al.}(2007)\citenamefont {Lee},
  \citenamefont {Zhang},\ and\ \citenamefont {Xiang}}]{lee2007edge}%
  \BibitemOpen
  \bibfield  {author} {\bibinfo {author} {\bibfnamefont {Dung-Hai}\
  \bibnamefont {Lee}}, \bibinfo {author} {\bibfnamefont {Guang-Ming}\
  \bibnamefont {Zhang}}, \ and\ \bibinfo {author} {\bibfnamefont {Tao}\
  \bibnamefont {Xiang}},\ }\bibfield  {title} {\enquote {\bibinfo {title} {Edge
  solitons of topological insulators and fractionalized quasiparticles in two
  dimensions},}\ }\href@noop {} {\bibfield  {journal} {\bibinfo  {journal}
  {Physical review letters}\ }\textbf {\bibinfo {volume} {99}},\ \bibinfo
  {pages} {196805} (\bibinfo {year} {2007})}\BibitemShut {NoStop}%
\bibitem [{\citenamefont {Shi}\ \emph {et~al.}(2009)\citenamefont {Shi},
  \citenamefont {Yu}, \citenamefont {You},\ and\ \citenamefont
  {Nori}}]{shi2009topological}%
  \BibitemOpen
  \bibfield  {author} {\bibinfo {author} {\bibfnamefont {Xiao-Feng}\
  \bibnamefont {Shi}}, \bibinfo {author} {\bibfnamefont {Yue}\ \bibnamefont
  {Yu}}, \bibinfo {author} {\bibfnamefont {JQ}~\bibnamefont {You}}, \ and\
  \bibinfo {author} {\bibfnamefont {Franco}\ \bibnamefont {Nori}},\ }\bibfield
  {title} {\enquote {\bibinfo {title} {Topological quantum phase transition in
  the extended kitaev spin model},}\ }\href@noop {} {\bibfield  {journal}
  {\bibinfo  {journal} {Physical Review B}\ }\textbf {\bibinfo {volume} {79}},\
  \bibinfo {pages} {134431} (\bibinfo {year} {2009})}\BibitemShut {NoStop}%
\bibitem [{\citenamefont {Ogura}\ \emph {et~al.}(2020)\citenamefont {Ogura},
  \citenamefont {Imamura}, \citenamefont {Kameyama}, \citenamefont {Minami},\
  and\ \citenamefont {Sato}}]{ogura2020geometric}%
  \BibitemOpen
  \bibfield  {author} {\bibinfo {author} {\bibfnamefont {Masahiro}\
  \bibnamefont {Ogura}}, \bibinfo {author} {\bibfnamefont {Yukihisa}\
  \bibnamefont {Imamura}}, \bibinfo {author} {\bibfnamefont {Naruhiko}\
  \bibnamefont {Kameyama}}, \bibinfo {author} {\bibfnamefont {Kazuhiko}\
  \bibnamefont {Minami}}, \ and\ \bibinfo {author} {\bibfnamefont {Masatoshi}\
  \bibnamefont {Sato}},\ }\bibfield  {title} {\enquote {\bibinfo {title}
  {Geometric criterion for solvability of lattice spin systems},}\ }\href@noop
  {} {\bibfield  {journal} {\bibinfo  {journal} {Physical Review B}\ }\textbf
  {\bibinfo {volume} {102}},\ \bibinfo {pages} {245118} (\bibinfo {year}
  {2020})}\BibitemShut {NoStop}%
\bibitem [{\citenamefont {Godsil}\ and\ \citenamefont {Royle}(2001)}]{godsil}%
  \BibitemOpen
  \bibfield  {author} {\bibinfo {author} {\bibfnamefont {Chris}\ \bibnamefont
  {Godsil}}\ and\ \bibinfo {author} {\bibfnamefont {Gordon}\ \bibnamefont
  {Royle}},\ }\href@noop {} {\emph {\bibinfo {title} {Algebraic Graph
  Theory}}}\ (\bibinfo  {publisher} {Springer},\ \bibinfo {year}
  {2001})\BibitemShut {NoStop}%
\bibitem [{\citenamefont {Bondy}\ and\ \citenamefont
  {Murty}(2008)}]{bondy2008graph}%
  \BibitemOpen
  \bibfield  {author} {\bibinfo {author} {\bibfnamefont {John~Adrian}\
  \bibnamefont {Bondy}}\ and\ \bibinfo {author} {\bibfnamefont {Uppaluri
  Siva~Ramachandra}\ \bibnamefont {Murty}},\ }\bibfield  {title} {\enquote
  {\bibinfo {title} {Graph theory (graduate texts in mathematics)},}\
  }\href@noop {} {\bibfield  {journal} {\bibinfo  {journal} {Grad. Texts in
  Math}\ }\textbf {\bibinfo {volume} {44}} (\bibinfo {year}
  {2008})}\BibitemShut {NoStop}%
\bibitem [{\citenamefont {Kozlov}(2008)}]{kozlov2008combinatorial}%
  \BibitemOpen
  \bibfield  {author} {\bibinfo {author} {\bibfnamefont {Dimitry}\ \bibnamefont
  {Kozlov}},\ }\href@noop {} {\emph {\bibinfo {title} {Combinatorial algebraic
  topology}}},\ Vol.~\bibinfo {volume} {21}\ (\bibinfo  {publisher} {Springer
  Science \& Business Media},\ \bibinfo {year} {2008})\BibitemShut {NoStop}%
\bibitem [{\citenamefont {Nakahara}(2003)}]{nakahara}%
  \BibitemOpen
  \bibfield  {author} {\bibinfo {author} {\bibfnamefont {Mikio}\ \bibnamefont
  {Nakahara}},\ }\href@noop {} {\emph {\bibinfo {title} {Geometry, Topology,
  and Physics}}}\ (\bibinfo  {publisher} {Taylor and Francis Group, LLC},\
  \bibinfo {year} {2003})\BibitemShut {NoStop}%
\bibitem [{Note1()}]{Note1}%
  \BibitemOpen
  \bibinfo {note} {We also present another theorem for the conserved
  quantities. See Ref.\cite {ogura2020geometric}.}\BibitemShut {Stop}%
\bibitem [{\citenamefont {Tsvelik}(2013)}]{tsvelik2013majorana}%
  \BibitemOpen
  \bibfield  {author} {\bibinfo {author} {\bibfnamefont {AM}~\bibnamefont
  {Tsvelik}},\ }\bibfield  {title} {\enquote {\bibinfo {title} {Majorana
  fermion realization of a two-channel kondo effect in a junction of three
  quantum ising chains},}\ }\href@noop {} {\bibfield  {journal} {\bibinfo
  {journal} {Physical review letters}\ }\textbf {\bibinfo {volume} {110}},\
  \bibinfo {pages} {147202} (\bibinfo {year} {2013})}\BibitemShut {NoStop}%
\bibitem [{\citenamefont {Giuliano}\ \emph {et~al.}(2016)\citenamefont
  {Giuliano}, \citenamefont {Campagnano},\ and\ \citenamefont
  {Tagliacozzo}}]{giuliano2016junction}%
  \BibitemOpen
  \bibfield  {author} {\bibinfo {author} {\bibfnamefont {Domenico}\
  \bibnamefont {Giuliano}}, \bibinfo {author} {\bibfnamefont {Gabriele}\
  \bibnamefont {Campagnano}}, \ and\ \bibinfo {author} {\bibfnamefont {Arturo}\
  \bibnamefont {Tagliacozzo}},\ }\bibfield  {title} {\enquote {\bibinfo {title}
  {Junction of three off-critical quantum ising chains and two-channel kondo
  effect in a superconductor},}\ }\href@noop {} {\bibfield  {journal} {\bibinfo
   {journal} {The European Physical Journal B}\ }\textbf {\bibinfo {volume}
  {89}},\ \bibinfo {pages} {251} (\bibinfo {year} {2016})}\BibitemShut
  {NoStop}%
\bibitem [{\citenamefont {Giuliano}\ \emph {et~al.}(2020)\citenamefont
  {Giuliano}, \citenamefont {Trombettoni},\ and\ \citenamefont
  {Sodano}}]{giuliano2020emerging}%
  \BibitemOpen
  \bibfield  {author} {\bibinfo {author} {\bibfnamefont {Domenico}\
  \bibnamefont {Giuliano}}, \bibinfo {author} {\bibfnamefont {Andrea}\
  \bibnamefont {Trombettoni}}, \ and\ \bibinfo {author} {\bibfnamefont
  {Pasquale}\ \bibnamefont {Sodano}},\ }\bibfield  {title} {\enquote {\bibinfo
  {title} {Emerging majorana modes in junctions of one-dimensional spin
  systems},}\ }\href@noop {} {\bibfield  {journal} {\bibinfo  {journal} {arXiv
  preprint arXiv:2002.06677}\ } (\bibinfo {year} {2020})}\BibitemShut {NoStop}%
\bibitem [{\citenamefont {Lieb}(2004)}]{lieb2004flux}%
  \BibitemOpen
  \bibfield  {author} {\bibinfo {author} {\bibfnamefont {Elliott~H}\
  \bibnamefont {Lieb}},\ }\bibfield  {title} {\enquote {\bibinfo {title} {Flux
  phase of the half-filled band},}\ }in\ \href@noop {} {\emph {\bibinfo
  {booktitle} {Condensed Matter Physics and Exactly Soluble Models}}}\
  (\bibinfo  {publisher} {Springer},\ \bibinfo {year} {2004})\ pp.\ \bibinfo
  {pages} {79--82}\BibitemShut {NoStop}%
\bibitem [{\citenamefont {Macris}\ and\ \citenamefont
  {Nachtergaele}(1996)}]{macris1996flux}%
  \BibitemOpen
  \bibfield  {author} {\bibinfo {author} {\bibfnamefont {Nicolas}\ \bibnamefont
  {Macris}}\ and\ \bibinfo {author} {\bibfnamefont {Bruno}\ \bibnamefont
  {Nachtergaele}},\ }\bibfield  {title} {\enquote {\bibinfo {title} {On the
  flux phase conjecture at half-filling: an improved proof},}\ }\href@noop {}
  {\bibfield  {journal} {\bibinfo  {journal} {Journal of statistical physics}\
  }\textbf {\bibinfo {volume} {85}},\ \bibinfo {pages} {745--761} (\bibinfo
  {year} {1996})}\BibitemShut {NoStop}%
\end{thebibliography}%

\end{document}